\documentclass[twocolumn]{aastex62}

\newcommand{\beq}{\begin{equation}}
\newcommand{\eeq}{\end{equation}}
\newcommand{\beqa}{\begin{eqnarray}}
\newcommand{\eeqa}{\end{eqnarray}}
\newcommand{\ter}{Terzan~5 }
\newcommand{\terc}{Terzan~5, }
\newcommand{\terp}{Terzan~5. }
\newcommand{\dgr}{$^\circ$}
\accepted{\today}
\submitjournal{ApJ}

\shorttitle{Multi-wavelength studies of Terzan~5}
\shortauthors{Ndiyavala et al.}

\begin{document}

\title{PROBING THE PULSAR POPULATION OF TERZAN~5 VIA SPECTRAL MODELING}


\collaboration{Fermi Large Area Telescope (LAT)}
\author[0000-0001-9279-1775]{H. NDIYAVALA}
\affiliation{Centre for Space Research, North-West University, Potchefstroom Campus, Private Bag X6001, Potchefstroom 2520, South Africa}
\affiliation{University of Namibia, Khomasdal Campus, Private Bag 13301, Windhoek, Namibia}

\author[0000-0002-2666-4812]{C. VENTER}
\affiliation{Centre for Space Research, North-West University, Potchefstroom Campus, Private Bag X6001, Potchefstroom 2520, South Africa}

\author{T.~J. JOHNSON}
\affiliation{College of Science, George Mason University, Fairfax, VA 22030, resident at Naval Research Laboratory, Washington, DC 20375, USA}

\author[0000-0001-6119-859X]{A.~K. HARDING}
\affiliation{Astrophysics Science Division, NASA Goddard Space Flight Center, Greenbelt, MD 20771, USA}

\author[0000-0002-7833-0275]{D.~A.~SMITH}  
\affiliation{Centre d'\'Etudes Nucl\'eaires de Bordeaux Gradignan, IN2P3/CNRS, Universit\'e Bordeaux 1, BP120, F-33175 Gradignan Cedex, France}

\nocollaboration
\author{P. EGER}
\affiliation{Universit\"at Erlangen-N\"urnberg, Physikalisches Institut, Erwin-Rommel-Str. 1, 91058 Erlangen, Germany}

\author[0000-0001-5206-9609]{A. KOPP}
\affiliation{Centre for Space Research, North-West University, Potchefstroom Campus, Private Bag X6001, Potchefstroom 2520, South Africa}
\affiliation{Institut f\"ur Theoretische Physik IV, Ruhr-Universit\"at Bochum, D-44780 Bochum, Germany}

\author[0000-0001-6666-6866]{D.~J. VAN DER WALT}
\affiliation{Centre for Space Research, North-West University, Potchefstroom Campus, Private Bag X6001, Potchefstroom 2520, South Africa}

\begin{abstract}
\ter is the only Galactic globular cluster that has plausibly been detected at very-high energies by the High Energy Stereoscopic System. It has an unexpectedly asymmetric very-high-energy morphology that is offset from the cluster center, in addition to a large-scale, offset radio structure and compact diffuse X-ray emission associated with this cluster. We present new data from the \textit{Fermi} Large Area Telescope on this source. We model the updated broadband spectral energy distribution, attributing this to cumulative pulsed emission from a population of embedded millisecond pulsars as well as unpulsed emission from the interaction of their leptonic winds with the ambient magnetic and soft-photon fields. In particular, our model invokes unpulsed synchrotron and inverse Compton components to model the radio and TeV data, cumulative pulsed curvature radiation to fit the \textit{Fermi} data, and explains the hard \textit{Chandra} X-ray spectrum via a ``new'' cumulative synchrotron component from electron-positron pairs within the pulsar magnetospheres that has not been implemented before. We find reasonable spectral fits for plausible model parameters. We also derive constraints on the millisecond pulsar luminosity function using the diffuse X-ray data and the \textit{Chandra} sensitivity. Future higher-quality spectral and spatial data will help discriminate between competing scenarios (such as dark matter annihilation, white dwarf winds or hadronic interactions) proposed for the broadband emission as well as constrain degenerate model parameters.
\end{abstract}


\section{INTRODUCTION}
Discovered in the 1960s, the Galactic globular cluster (GC) \ter is a fascinating object lying at a distance $d=5.9\pm0.5$~kpc \citep{Valenti07} and having a particularly high central stellar density as well as high metallicity. It also has the highest stellar interaction rate of all Galactic GCs \citep{Verbunt87}, which is probably linked to the large number of X-ray binaries found in this system. The latter may furthermore explain the fact that \ter hosts the largest number of millisecond pulsars ($N^{\rm rad}_{\rm vis}=37$ visible radio MSPs) of all Galactic GCs \citep{Cadelano18},
MSPs being the offspring of low-mass X-ray binaries \citep{Camilo05,Abdo10}. The discovery of two distinct stellar populations with different iron content and ages in this GC has been interpreted as an indication that \ter may not be a ``true'' GC in the usual sense: it may represent the merger of two stellar clusters, or it may be the remnant of a disrupted galaxy \citep{Ferraro09}. The high metallicity probably points to a very large number of supernova explosions (i.e., progenitors of neutron stars) occurring in \terc further explaining why this cluster harbors so many MSPs. Moreover, the latter leads to the expectation that even more MSPs may be found in \ter than the current 37 known ones\footnote{http://www.naic.edu} \citep{Lanzoni10, Freire17,Cadelano18}. As such, this GC has been an attractive source to model and observe, since MSPs are known not only to radiate pulsed emission in multiple wavebands, but to generate relativistic particles that may in turn upscatter ambient photons into the very-high-energy (VHE) domain \citep{BS07}, or interact with the cluster magnetic field to yield diffuse synchrotron radiation (SR; \citealt{Venter08}). 
 
The \textit{Fermi} Large Area Telescope \citep[LAT;][]{Atwood09} has detected a bright GeV source that is very plausibly associated with \ter \citep{Abdo10,2FGL}, bringing the total number of LAT sources that are associated with GCs up to about 20 \citep{3FGL,Zhang16}. \citet{deMenezes} recently performed a systematic study of 23 \textit{Fermi} GC candidates and detected \ter at more than 60$\sigma$. The High Energy Stereoscopic System (H.E.S.S.) has furthermore detected an extended source in the direction of \terc although its morphology is peculiar and offset from the GC center \citep{Abramowski11}. This makes \ter the only GC plausibly detected at VHEs \citep{Anderhub09,Aharonian09,McCutcheon09,Abramowski11_DM,Abramowski13}. (See \citealt{Tam16} for a recent review of $\gamma$-ray detections of GCs.) Diffuse X-rays have also been detected from this GC, peaking at the center and decreasing with cluster radius \citep{Eger10}. Radio observations have revealed several extended structures in the vicinity of this source \citep{Clapson2011}. In light of the available multi-wavelength data and the unique and unexpected source morphology in different energy bands, this source represents a prime subject for deeper investigation. 

Several models have attempted to explain the broadband emission properties of GCs. The first class of models invoke MSPs as sources of relativistic particles and cumulative high-energy emission. \citet{Chen1991} provided an early estimate of the cumulative $\gamma$-ray luminosity from a population of MSPs embedded in a GC, finding $L_{\rm \gamma,tot}\sim10^{36}n_{500}$~erg\,s$^{-1}$ for $n_{500}=N^\gamma_{\rm MSP}/500$ and $N^\gamma_{\rm MSP}$ the number of $\gamma$-ray-bright MSPs, when convolving the predicted $L_\gamma$ for each pulsar with an expected distribution of periods of GC MSPs (see also \citealt{Bhatia92}). This estimate turns out to be correct to within a factor of a few compared to the measured GeV luminosities for some \textit{Fermi}-detected GCs \citep[e.g.,][]{Abdo10} if one sets $n_{500}\sim0.2$. \citet{Wei96} calculated such a cumulative $\gamma$-ray flux using an outer gap model. Comparing their expected flux level for unpulsed $\gamma$-rays 
to an upper limit by the Energetic Gamma-Ray Experiment Telescope (\textit{EGRET}), they constrained $n_{500}<0.8$ for 47~Tucanae. \citet{HUM05,Venter08,Venter09_GC} summed the individual pulsed curvature radiation (CR) spectra from their model for an ensemble of MSPs to estimate the GeV flux expected from a GC, and the predictions of \citet{Venter09_GC} provided a good match to the subsequent {\it Fermi} measurements of the high-energy (HE) spectrum of 47~Tucanae~\citep{Abdo09}. \citet{Cheng10} investigated an alternative scenario to produce GeV emission by attributing this to inverse Compton (IC) rather than CR emission, also predicting GCs to be extended sources in the GeV regime. Conversely, \citet{BS07} predicted that GCs may be point-like sources of GeV and TeV emission by considering MSPs that accelerate leptons either at the shocks that originate during collisions of the respective pulsar winds, or inside the pulsar magnetospheres. The leptons escape from these local acceleration sites, diffuse outward, and interact with the GC magnetic and soft-photon background fields. This leads to SR and IC scattering (see \citealt{Venter09_GC} and \citealt{Zajczyk13} for updated calculations). \citet{Kopp13} presented an improved model and found reasonable fits to the multi-band spectral energy distribution (SED) data of \terp \citet{Ndiyavala18} applied this model to the Galactic population of GCs, ranking them according to predicted VHE flux.

There exist alternative models that invoke other astrophysical objects as sources of relativistic particles. \citet{Bednarek12} calculated the contribution of non-accreting white dwarfs in GCs to the $\gamma$-ray flux from such clusters and concluded that white dwarfs may produce $\gamma$-ray emission at a level which may be detectable by the Cherenkov Telescope Array (CTA) in some cases. See \citet{Bednarek11} for a review of the such leptonic GC models. On the other hand, \citet{Domainko11} investigated a hadronic model, invoking a $\gamma$-ray burst remnant as a potential source of energetic leptons and hadrons. In this model, the hadrons interact with ambient target nuclei, leading to the formation of $\pi^0$ particles that decay into $\gamma$ rays. Recently, \citet{Brown18} concluded\footnote{These authors themselves note that their work does not rule out an MSP-only explanation for the GeV flux seen from 47~Tucanae. They used the average spectrum of \textit{Fermi}-detected MSPs, assuming this to be universal, and allowing only the spectral normalization to be free. Other effects such as the inclusion of MSPs that are below detection threshold as well as different MSP geometries (inclination and viewing angles) may change the low-energy spectral shape, hardening it to potentially bring it in better agreement with the data, without the need to invoke dark matter annihilation (the latter model has several more free parameters, and a combination of MSPs and dark matter may therefore naturally better account for the data).} that a combination of MSP pulsed curvature emission and dark matter annihilation (with an enhanced density around a putative intermediate-mass black hole)  may explain the GeV emission detected by \textit{Fermi} LAT for 47~Tucanae.

Even though models are making progress to explain the broadband emission of GCs, many questions remain. For example, \citet{Venter_HEASA15} noted that uncertainties in the model parameters may lead to a spread in the predicted GC flux of up to an order of magnitude. They attempted to mitigate this problem by considering an ensemble of observed GCs to constrain their models, using a H.E.S.S.\ upper limit \citep{Abramowski13} to the cumulative flux from 15 GCs (\citealt{Venter15,Venter_HEASA15}, Ndiyavala et al., in prep.), but some parameter degeneracies are expected to remain. The hard slope of the diffuse X-ray emission in the case of \ter poses another puzzle, since the existing models have not been able to fit this component \citep[e.g.,][]{Kopp13}. The energy-dependent morphology (which is non-spherical at high energies) further challenges the existing models. \citet{Bednarek14} considered a model where energetic particles escape from the GC and interact with the Galactic medium, creating a bow shock nebula around the GC. If the latter is immersed in the relatively dense medium close to the Galactic Plane, this should manifest as an intricate morphology at high energies. To further address this complex morphology, \citet{Bednarek16} extended their model to take into account the advection of leptons by a mixture of red giant stellar and pulsar winds, as well as considering the effect of having a non-central (offset) energetic MSP as a source of relativistic particles. Furthermore, in the case of \terc the source morphologies differ significantly in extent and position across the electromagnetic spectrum, raising the question whether all the spectral components arise due to the same underlying particle population (in the leptonic scenario) or not. Lastly, the operation of different emission mechanisms and relative contribution of MSPs vs.\ other astrophysical sources or dark matter to the SED remains an open question.

Given the richness of the existing data set on \terc as well as the variety of models that exist to explain GC emission (and their many free parameters), we use this system as a case study to further probe the origin of multi-wavelength emission from GCs. Improved models will aid selection of promising GCs for future observations by the CTA, which may see tens of these sources in the next decade \citep{Ndiyavala18}. We therefore aimed to gather more data on \ter (Section~\ref{sec:data}) and model the updated SED in a leptonic scenario (Section~\ref{sec:leptonic}). We present our conclusions in Section~\ref{sec:concl}.

\section{MULTI-WAVELENGTH DATA AND SPECTRAL UPPER LIMITS}\label{sec:data}

\subsection{Previous Radio Observations}
Individual MSP discoveries in \ter bring the total membership to 37 \citep[e.g.,][]{Lyne90,Lyne00,Ransom05,Hessels06,Freire16, Freire17,Cadelano18}, although hundreds of MSPs may be present following expectations of numerical simulations \citep{Ivanova08}. \citet{Fruchter00} obtained images of \ter at 6~cm, 20~cm, and 90~cm using the Very Large Array (VLA). These displayed strong, steep-spectrum emission that could not be associated with known pulsars at that time. Numerous point sources were also detected within $30^{\prime\prime}$ of the cluster center, with their density rising rapidly toward the core. There, an elongated region of emission was found. Based on the steep spectrum as well as on the flux distribution, \citet{Fruchter00} concluded that this most probably indicated the presence of many undetected pulsars in the cluster, making this the most pulsar-rich of all Galactic GCs (they estimated a total number of $60-200$ host pulsars, based on their assumed radio luminosity function). 

\ter was furthermore detected in the NRAO VLA Sky Survey (NVSS) at 21~cm as a single source with a flux of about 5~mJy \citep{Condon98}. \citet{Clapson2011} analyzed archival 11~cm and 21~cm Effelsberg data and detected several radio structures in the direction of \terp However, given the uncertainty in flux, no spectral index could be inferred. \citet{Clapson2011} speculated that one structure\footnote{Care should be taken to simply compare the results of \citet{Condon98} with those of \citet{Clapson2011}. The NVSS was done with the VLA in the DnC configuration and the largest angular scale that can be detected in this most compact configuration is 970$^{\prime\prime}$ in full synthesis mode. In snapshot mode, this is 485$^{\prime\prime}$. We expect that only the brightest part of the emission would have been detected and that the angular size of the detected region was actually less than 485$^{\prime\prime}$.  In fact, \citet{Condon98} give the size of the emission region as less than $58^{\prime\prime}\times47^{\prime\prime}$. Conversely, \citet{Clapson2011} used the Effelsberg single-dish telescope for which there is no limit to the largest detectable angular size of an extended structure. In the present context, the largest angular scale is of importance. \citet{Clapson2011} listed a size for Region 11 of $720^{\prime\prime}\times1080^{\prime\prime}$, even larger than the tidal radius of \ter of $R_{\rm t}\sim280^{\prime\prime}$, and also exceeding the largest angular size detectable at 21~cm with the VLA in the DnC configuration. It thus makes sense that the implied flux measured by \citet{Clapson2011} for Region 11 at 21~cm is $\sim4$~Jy vs.\ the relatively low flux of $\sim5$~mJy measured by \citet{Condon98}. However, the surface brightness of both observations are quite similar in magnitude. It is unquestionable that the \citet{Condon98} observations suffered from the ``missing-flux effect'' of interferometric observations. We conclude that the \citet{Clapson2011} value of the total flux from Region 11 is the more reliable value that should be used in the model fitting.} in particular (labeled as ``Region~11''), extending from the GC center to the north-west (roughly perpendicular to the Galactic Plane), could be the result of SR by electrons escaping from the large population of MSPs in this GC. In what follows, we fit these radio data using a diffuse low-energy synchrotron radiation (LESR) component due to the interaction of relativistic electrons escaping from the MSP magnetospheres with the cluster $B$-field (see Figure~\ref{fig:SED}). The contribution of the  population of unresolved MSPs to this diffuse radio flux is negligible and can be ignored\footnote{\citet{Ransom05} estimated the total flux at 1.95~GHz of 22 MSPs in the core of \ter to be $\sim1$~mJy (the scale is $R_{\rm c}\sim9^{\prime\prime}$) or up to a few mJy if one adds two more MSPs farther from the GC center, while the flux from the large Region~11 measured by \citet{Clapson2011} of $\sim4$~Jy dwarfs this value.}.

\subsection{Optical Upper Limits: Comparison of Thermal and non-Thermal Flux Levels}
\label{optical}
Our predicted non-thermal unpulsed LESR spectral component invoked to model the radio data (Section~\ref{sec:model}) extends into the optical band, raising the question of its detectability. The LESR component's flux is relatively low in the optical band \citep{Kopp13}, and we now show that it may in principle be very difficult to directly observe it, since there are $\sim 10^5$ stars (point sources) that contribute a high level of blackbody (BB) radiation that will swamp any diffuse non-thermal SR. 

To appraise the BB $\nu F_\nu$ flux from stars in different annuli centered on the cluster, and also the total flux expected from the full cluster, we use the surface-density profile of \ter obtained by \citet{Trager1995}, as converted by \citet{Cohn2002}. We estimate the area of an annulus as $ A_{\rm ann} = \pi(\theta_{2}^{2} - \theta_{1}^{2})$, with $\theta_{1}$ and $\theta_{2}$ the edges (angular radii) of a particular annulus. The average number of stars in such an annulus is found by interpolating the surface brightness $f$ and calculating $ N_{\rm ann} =  fA_{\rm ann}$. We approximate the emitted spectrum of each star by a BB spectrum at a single average frequency $\langle\nu\rangle=2.7k_{\rm B}T/h=2.5\times10^{14}$~Hz, where $k_{\rm B}$ is the Boltzmann constant and $h$ the Planck constant, assuming a constant stellar surface temperature of $T = 4~500\,\rm K$. Upon multiplying the Planck spectrum $B_{\nu}$ by the stellar surface area $A_{\star}=4\pi R_*^2$, with $R_*$ the average stellar radius, and dividing by the square of the distance to the cluster, we obtain the thermal $\nu F_{\nu}$ flux level:
\begin{eqnarray}
 \frac{B_{\nu}\langle\nu\rangle A_{\star}}{d^{2}} & = & \frac{8\pi R_*^{2}h\langle\nu\rangle^{4}N_{\rm ann}}{d^{2}c^{2}}\frac{1}{e^{h\nu/k_{B}T}-1} \nonumber \\
& \sim & 1.7\times10^{-14}R^{2}_{*,10}N_{\rm ann}\,\rm erg\,cm^{-2}s^{-1},\label{eq:BB_flux} 
\end{eqnarray}
with $R_{*,10} = R_*/10^{10}\,\rm cm$, $c$ the speed of light, and assuming $d\,=\,5.9~{\rm kpc}$. 
When applying Eq.~(\ref{eq:BB_flux}) to the whole cluster (i.e., choosing $N_*=N_{\rm ann} = 7.7\times10^4$; \citealt{Lang1992}), we find that the predicted BB flux is $\sim 6.2\times10^{-8}$~erg\,cm$^{-2}$s$^{-1}$ for $R_{10}=7$, while the predicted $\nu F_\nu$ flux for the LESR at 1~eV is only $\sim5.0\times10^{-12}$~erg\,cm$^{-2}$s$^{-1}$ (Section~\ref{sec:model}), which is a factor $\sim10^4$ lower than the estimated thermal flux level. The only hope to detect the LESR component is if the stellar flux falls faster with radius than does the LESR flux. One might thus try to obtain a smaller ratio between thermal and non-thermal fluxes by focusing on different annuli. We calculated the flux ratio for all annuli in \citet{Cohn2002} and found that they drop from $\sim10^5$ to $\sim10^3$ with increasing radius, out to $\sim0.35R_{\rm t}$, with $R_{\rm t}$ the tidal radius. Since the annular area increases as $r^2$ while the surface brightness falls as $r^{-2.1}$ for larger radii, the estimated BB flux remains nearly constant for the outer annuli (with $N_{\rm ann}\sim10^3$) out to $R_{\rm t}$. Thus, even in the outer annuli where the stellar density has dropped significantly, the BB flux still exceeds the LESR flux by a factor $10^3$. In some sense, the thermal emission thus provides a very unconstraining upper limit to the LESR emission under the assumption that the latter will not exceed the thermal flux. However, since the BB spectrum covers a much narrower energy range than the LESR, there may yet be hope of detecting the LESR flux outside of the optical range (e.g., the millimeter or ultraviolet to low-X-ray range) where the BB component dominates.

\subsection{Diffuse X-ray Emission}
To investigate the X-ray point source population of \terc the core of this GC was covered in a deep \textit{Chandra} observation with a field of view of $\sim 5^\prime\times 5^\prime$ \citep{Heinke06}. Later, by following up on the detection of extended TeV $\gamma$-ray emission from the direction of \ter by H.E.S.S.\ \citep{Abramowski11}, \citet{Eger10} discovered the presence of hard and diffuse X-ray emission using the same \textit{Chandra} data. The diffuse X-ray signal was shown to be extended well beyond the half-mass-radius ($R_{\rm hm}\sim30^{\prime\prime}$) of the GC up to $\sim180^{\prime\prime}$, featuring a very hard spectrum that may be fit by a power law with a photon index of $0.9\pm0.5$ (see Figure~\ref{fig:SED}).
The contribution from unresolved point-like sources to this diffuse signal was estimated to be very small. 
They found that the surface brightness peaked near the cluster center and decreased smoothly outwards. Various non-thermal emission mechanisms for the origin of this diffuse signal were discussed, but with no single scenario being clearly preferred. A follow-up search for an X-ray signal on similar spatial scales from a number of other LAT-detected GCs covered by archival X-ray observations yielded no additional significant detections \citep[see][]{Eger12}. However, a new hard, diffuse X-ray signal was recently discovered from 47~Tucanae, yet on comparatively smaller spatial scales \citep{Wu14}. In contrast to \terc the X-ray signal here appears to be contained within the half-mass radius of the GC. The spectrum can be described as a combination of a hard power-law component with a photon index of $\sim$1.0, and a thermal plasma component with a temperature of $k_{\rm B}T = 0.2$\,keV. The non-thermal X-ray emission detected from both \ter and 47~Tucanae could be unpulsed SR from relativistic leptons that were accelerated in shocks, following the collision of stellar winds in the GC cores (i.e., a single spectral component explaining both the radio and X-ray data in the case of \terc although the diffuse X-ray emission appears on very different spatial scales in these two GCs; \citealt{BS07,Venter09_GC}). However, we cannot find a satisfactory fit to the spectral data of these clusters when invoking only a single SR spectral component. We therefore model the diffuse X-ray emission observed from \ter by invoking a new component that is due to the cumulative pulsed SR by pairs originating in the various host MSP magnetospheres (Section~\ref{sec:model}).

\subsection{New \textit{Fermi} LAT Data Analysis}\label{sec:latp8}
\ter was the second GC to be associated with a \textit{Fermi}-LAT source \citep{Kong10,Abdo10,2FGL}. Comparing the likelihood when modeling the spectrum with a simple power-law shape,
\beq\label{eq:pl}
\frac{dN}{dE}\ =\ N_{0}\ \left(\frac{E}{E_{0}}\right)^{-\Gamma}
\eeq
\noindent{}and an exponentially cutoff power-law shape:
\beq\label{eq:ecpl}
\frac{dN}{dE}\ =\ N_{0}\ \left(\frac{E}{E_{0}}\right)^{-\Gamma}\ \exp\left\{-\left(\frac{E}{E_{\rm C}} \right)^{b} \right\}
\eeq
\noindent{}the $\gamma$-ray point source associated with \ter was found to be significantly curved, consistent with the interpretation of the collective emission from a population of MSPs.  In both Eq.~(\ref{eq:pl}) and (\ref{eq:ecpl}), $N_{0}$ is a normalization factor with units cm$^{-2}$\,s$^{-1}$\,MeV$^{-1}$, $E_{0}$ is a scale parameter, and $\Gamma$ is the photon index. In Eq.~(\ref{eq:ecpl}), $E_{\rm C}$ is the cutoff energy and $b$ is an exponential index that governs how quickly the spectrum rolls over. Low-altitude pulsar emission models predict a super-exponential cutoff with $b\ >\ 1$ \citep[e.g.,][]{Harding78}. For some of the brightest $\gamma$-ray pulsars, LAT observations require a sub-exponential cutoff with $b\ <\ 1$, plausibly explained as a blending of several simple exponentially-cutoff spectra as the line of sight crosses different regions of the magnetosphere \citep{2PC}.

\citet{Kong10} analyzed approximately 1.4 years of Pass 6 (P6) LAT data from the region around \ter with energies ranging from 0.5 to 20 GeV, and found a significant point source (18$^\prime$) from the optical center of \terc with a pulsar-like spectrum having a photon index $\Gamma\ =\ 1.9\pm0.2$, a cutoff energy $E_{\rm C}\ =\ 3.8\pm 1.2$~GeV, and integrated photon and energy fluxes over their energy range of $(3.4\pm1.1)\times 10^{-8}$ cm$^{-2}$\,s$^{-1}$ and $(6.8\pm2.0)\times 10^{-11}$ erg\,cm$^{-2}$\,s$^{-1}$, respectively.

\citet{Abdo10} analyzed the region around \ter using approximately 1.5 years of P6 LAT data, with energies $\geq$ 0.2 GeV, including the same time span of \citet{Kong10}. These authors also found a significant point source, located $2.4^\prime$ from the cluster center, with a pulsar-like spectrum. Their best-fit simple exponentially-cutoff power-law spectrum had $\Gamma\ =\ 1.4^{+0.2,+0.4}_{-0.2,-0.3}$, $E_{\rm C}\ =\ 2.6^{+0.7,+1.2}_{-0.5,-0.7}$ GeV, and integrated photon and energy fluxes over their energy range of $(7.6^{+1.7,+3.4}_{-1.5,-2.2})\times 10^{-8}$ cm$^{-2}$\,s$^{-1}$ and $(7.1^{+0.6,+1.0}_{-0.5,-0.5})\times 10^{-11}$ erg cm$^{-2}$\,s$^{-1}$, respectively.  The first uncertainties are statistical while the second reflect estimates of systematic errors.  Using an estimate of the average MSP spin-down power and $\gamma$-ray efficiency with the measured $\gamma$-ray flux, \citet{Abdo10} estimated the number of MSPs in \ter to be $N^\gamma_{\rm MSP}\ =\ 180^{+100}_{-90}$.

Using four years of data, the third \emph{Fermi} LAT catalog \citep[3FGL,][]{3FGL} associates 3FGL J1748.0$-$2447 with \terp The source is offset from the cluster center\footnote{The optical and diffuse X-ray centers are more or less coincident. The center of the radio ``Region 11'' is offset from this position by $\sim14^\prime$, while that of the extended H.E.S.S.\ source is offset by $\sim4^\prime$ (compared to a tidal radius of $R_{\rm t}=4.6{^\prime}$).} by $0.66^\prime$, well within the 95\% confidence-level ellipse with semi-major and semi-minor axes of $1.69^\prime$ and $1.53^\prime$, respectively.   The pivot energy for 3FGL J1748.0$-$2447, 1280.38 MeV, is used as the scale parameter $E_{0}$ in our subsequent analyses.

We selected seven years of Pass~8 (P8) LAT data\footnote{\url{http://fermi.gsfc.nasa.gov/ssc/data/analysis/}\\\url{documentation/Pass8_usage.html}} \citep{AtwoodP8} from the start of science operations on 2008 August 4, with \textit{evclass} = 128 and \textit{evtype} = 3, within 15\degr\ of the best-fit position of 3FGL J1748.0$-$2447, with energies from 0.1 to 300 GeV, and with maximum zenith angle of 90\dgr.  The \textit{Fermi} ScienceTool\footnote{Available for download at \url{https://fermi.gsfc.nasa}} (ST) \texttt{gtmktime} was used to select good time intervals when the spacecraft was in nominal science operations mode and the data were flagged as good.  In preparation for a binned maximum likelihood analysis, we made a livetime cube using the ST \texttt{gtltcube} with \textit{zmax} = 90\dgr\ and an exposure cube with 35 bins in log$_{10}$ energy and spatial pixels 0\fdg1 on a side using the ST \texttt{gtexpcube2} and the P8R2\_SOURCE\_V6 LAT Instrument Response Functions.

We constructed a model of our region of interest (ROI) including all 3FGL sources within 25\dgr\ of the ROI center, those sources known to be extended were modeled using the spatial templates from the catalog. The spectral parameters of sources $>$ 6\dgr\ from the ROI center were held fixed at the values from 3FGL. For sources within 6\dgr\ of the ROI center, the spectral parameters were allowed to vary if they were found to have an average significance $\geq$ 15$\sigma$ in 3FGL.  However, for sources within 8\dgr\ of the ROI center that did not otherwise satisfy our requirements for free spectral parameters but were flagged as significantly variable in 3FGL, we did allow the normalization parameters to vary.  The diffuse emission from the Milky Way was included using the \textit{gll\_iem\_v06.fits} model, while the isotropic diffuse emission and residual background of misclassified cosmic rays were modeled using the \textit{iso\_P8R2\_SOURCE\_V6\_v06.txt} template \citep{Acero16}. We allowed the intensity of the Galactic diffuse emission to be modified by a power-law spectrum.

The spectrum of 3FGL J1748.0$-$2447 was found to have significant curvature and, thus, modeled using a log-parabola function in the catalog.  For our purposes, we modeled the spectrum of 3FGL J1748.0$-$2447 using both a simple power law (Eq.~[\ref{eq:pl}]) and an exponentially-cutoff power law (Eq.~[\ref{eq:ecpl}]).  We performed three binned maximum likelihood analyses, with energy dispersion disabled, with the spectrum of 3FGL J1748.0$-$2447 modeled as a power law, as a simple exponentially-cutoff power law, and as an exponentially-cutoff power law with the $b$ parameter allowed to vary. Following \citet{2PC}, we compared the best-fit likelihood value from the fit using a power law ($\mathcal{L}_{\rm pl}$) to that when using a simple cutoff ($\mathcal{L}_{\rm co}$ when $b=1$) to calculate $TS_{\rm cut}$ = $-2(\ln(\mathcal{L}_{\rm co})-\ln(\mathcal{L}_{\rm pl}))$ = 207, significantly favoring the cutoff model over the power law. Similarly, we found $TS_{\rm{b free}}$ = $-2(\ln(\mathcal{L}_{\rm{b free}})-\ln(\mathcal{L}_{\rm co}))$ = 4, where $\mathcal{L}_{\rm{b free}}$ is the best-fit likelihood when modeling the spectrum of 3FGL J1748.0$-$2447 as an exponentially-cutoff power law with the $b$ parameter free. As such, there is no preference for the fit with $b$ free and we use the results from the simple exponentially-cutoff power law, which are reported in Table~\ref{tab:spec}. In additin to the fit parameters from Eq.~(\ref{eq:ecpl}), Table~\ref{tab:spec} also includes the integrated photon ($F$) and energy ($G$) fluxes, from 0.1 to 300 GeV, derived from the best-fit models.  Our best-fit energy flux agrees well with that of \citet{deMenezes}, who reported a value of $G_{100}$ = $(7.44\pm0.27)\times10^{-11}$ erg cm$^{-2}$ s$^{-1}$ (rounded so that there are two significant figures in the error), over the energy range 0.1 to 100 GeV, using nine years of P8 data, and assuming a log-parabola spectral shape.
A residual TS map of the region around 3FGL J1748.0$-$2447, using our best-fit $b=1$ model, did not reveal the need for adding any new sources to our model, even though the data sets we analyzed covered three more years than that used for the 3FGL catalog.

\begin{deluxetable}{lc}
\tablewidth{0pt}
\tablecaption{LAT Spectral Fit Results \label{tab:spec}}
\tablecolumns{2}
\startdata\\
$N_{0}$ (10$^{-11}$ cm$^{-2}$\,s$^{-1}$\,MeV$^{-1}$) & 1.04$\pm$0.40\\
$\Gamma$ & 1.71$\pm$0.04\\
$E_{\rm C}$ (GeV) & 4.61$\pm$0.35\\
$F_{100}$ (10$^{-8}$ cm$^{-2}$\,s$^{-1}$) & 9.68$\pm$0.52\\
$G_{100}$ (10$^{-11}$ erg\,cm$^{-2}$\,s$^{-1}$) & 7.79$\pm$0.22\\
$TS_{\rm cut}$ & 207\\
$TS_{\rm b free}$ & 4\\
\enddata
\end{deluxetable}

While our likelihood analyses did successfully converge, the fits of the entire region were formally bad.  In particular, there were large residuals starting at $\sim$10 GeV, growing to larger discrepancies out to 300 GeV.
The preliminary 8-year LAT source catalog\footnote{Available at \url{https://fermi.gsfc.nasa.gov/ssc/data/access/lat/8yr_catalog/}} with an improved Galactic diffuse model, P8R3 data, and using weighted likelihood \citep{4FGL} finds much better residuals in the region around Terzan~5.  This catalog used a different functional form for the spectrum of the source associated with Terzan~5 (4FGL J1748.0$-$2446) and fixed $b=2/3$, but we can still compare the flux values. The source 4FGL J1748.0$-$2446 has a reported integral photon flux, above 1 GeV, of (1.26$\pm$0.03)$\times10^{-8}$ cm$^{-2}$ s$^{-1}$, which is larger than the value of (1.05$\pm$0.03)$\times10^{-8}$ cm$^{-2}$ s$^{-1}$ found in our analysis.
Our best-fit $\Gamma$ value agrees with both previous studies \citep{Kong10, Abdo10}, within their quoted uncertainties.  The best-fit $E_{\rm C}$ of \citet{Kong10} agrees with our value, within uncertainty, but that of \citet{Abdo10} is significantly lower. Using our P8 results, we find photon and energy fluxes over the 0.5 to 20 GeV energy range of $(2.3\pm 0.1)\times 10^{-8}$ cm$^{-2}$\,s$^{-1}$ and $(5.3\pm 0.1)\times 10^{-11}$ erg cm$^{-2}$\,s$^{-1}$. While both values are lower than those reported by \citet{Kong10}, they agree within uncertainties.  Integrating above 0.2~GeV, our model yields photon and energy fluxes of $(5.4\pm 0.2)\times 10^{-8}$ cm$^{-2}$\,s$^{-1}$ and $(6.8\pm 0.2)\times 10^{-11}$ erg cm$^{-2}$\,s$^{-1}$. These values are also lower than those reported by \citet{Abdo10} but agree within uncertainties, and we note that the energy flux values \citep[often more reliable as noted by][]{2PC} agrees well.

We produced spectral points by dividing the 0.1 to 300 GeV interval into 12 bins, equally sized in log$_{10}$ energy, and performing binned likelihood fits assuming a power-law form for the spectrum of 3FGL J1748$-$2447 with $\Gamma = 2$ and only the normalization parameters of other sources left free.  We report a flux value, with uncertainty, for those bins where 3FGL J1748$-$2447 was detected with a point-source TS $\geq$ 9 ($\sim3\sigma$) and at least 4 predicted counts, otherwise a 95\% confidence-level flux upper limit is reported.  The flux upper limits were calculated using the Bayesian method for energy bins with point-source TS $\leq$ 0 or with $<$ 4 predicted counts from 3FGL J1748$-$2447.  For plotting and to produce the $E^{2}dN/dE$ points, we used the logarithmic mean of each energy bin (cf.\ Figure~\ref{fig:SED}). In Section~\ref{sec:model}, we model the \textit{Fermi} LAT spectrum as originating due to the cumulative pulsed CR by the embedded MSPs.

In order to search for $\gamma$-ray pulsations, we obtained timing solutions for 33 of the pulsars in \ter (namely, PSRs J1748$-$2446aa, ab, ac, ae, af, ag, ah, ai, aj, ak, C, D, E, F, G, H, I, J, K, L, M, N, O, Q, R, S, T, U, V, W, X, Y, and Z) that were valid from before the launch of \textit{Fermi} until 2012 July (S.\ Ransom, personal communication\footnote{\url{https://www.cv.nrao.edu/~sransom/Ter5_index.html}}).  We used the ephemerides for pulsars aj and ak from \citet{Cadelano18}. Using the best-fit maximum likelihood model, in which the spectrum of 3FGL J1748.0$-$2447 was modeled as a simple exponentially-cutoff power law, we calculated spectral weights for events within 2\degr\ of the best-fit position. For each event, the weight reflects the probability that the event originated from 3FGL J1748.0$-$2447. Use of these weights has been shown to enhance the sensitivity of $\gamma$-ray pulsation searches \citep{Kerr11}. We then used the timing solutions mentioned above to search for modulations in $\gamma$-ray events at the spin and orbital periods from known pulsars in \terp We tested for modulation at the spin period using the H test \citep{deJager89,deJager10}, modified to include spectral weights \citep{Kerr11}. For those pulsars in binary systems we used both the H test and the $Z^{2}_{m}$ test with $m$ = 2 harmonics when testing for modulation at the orbital period. When performing the search for orbital modulation, we corrected for exposure variations as described in \citet{JohnsonJ1227}. We tested for spin pulsations using both the full data set and only events up to the end of each ephemeris' validity interval.  No significant modulation was detected from any pulsar for which we had a timing solution, with a maximum signal of 2.2$\sigma$.

\subsection{H.E.S.S.\ Data}
\label{sec:hess}
H.E.S.S.\ discovered a VHE $\gamma$-ray source in the direction of \ter \citep{Abramowski11}. The integral flux above 440~GeV of the source was measured as $(1.2 \pm 0.3) \times 10^{-12}$~cm$^{-2}$\,s$^{-1}$ and its spectrum was best described by a single power law with index of $2.5 \pm 0.3_\mathrm{stat} \pm 0.2_\mathrm{sys}$. The VHE source is offset from the center of the GC by $4.0^\prime\pm1.9^\prime$ (about 7~pc at a distance of 5.9~kpc), with its size being characterized by widths from a 2D Gaussian fit of $9.6^\prime\pm2.4^\prime$ and $1.8^\prime\pm1.2^\prime$ for the major and minor axes (compared to the GC tidal radius of $R_{\rm t}=4.6^\prime$). The sources is oriented 92\dgr$\pm$6\dgr\ westwards from north\footnote{This means that the H.E.S.S.\ source is much closer to the GC core than the radio ``Region~11'', and it only slightly overlaps with its inner edge.}. A chance coincidence between \ter and an unrelated VHE $\gamma$-ray source is rather unlikely ($\sim 10^{-4}$). \citet{Ndiyavala18} reanalyzed \ter data and obtained a significance of $6\sigma$ for standard and loose cuts, and $7.1\sigma$  for hard cuts that compare well with that of \citet{Abramowski11} who obtained a significance of $5.3\sigma$.

\section{MODELING THE BROADBAND SED}\label{sec:model}

\subsection{Parameter Constraints from General Considerations}
In this section, we derive general constraints on the spatial diffusion coefficient $\kappa$ (for simplicity, we assume that this coefficient is only a function of particle energy, not of space) and the cluster $B$-field. 

As a first approach, the Bohm value has been used in the past to model the particle diffusion \citep[][]{BS07,Venter08}:
\beq
\kappa_{\rm Bohm} = \frac{cE_{\rm e}}{3eB} = 3.3\times10^{25}E_{\rm TeV}B_{-6}^{-1}~{\rm cm}^2\,{\rm s}^{-1},\label{eq:Bohm}
\eeq
with $E_{\rm TeV}=E_{\rm e}/{\rm 1~TeV}$ the particle energy and $B_{-6}=B/1~\mu{\rm G}$.
By invoking a containment argument, one may obtain a constraint on this coefficient: since we observe VHE $\gamma$-ray emission up to $E_\gamma\sim10$~TeV, one can write that the diffusion time (in the limit that it exceeds the escape time) should exceed the typical timescale for IC emission:
\beq
\tau_{\rm esc} > \tau_{\rm IC}.
\eeq
This leads to
\beq
\frac{R^2}{6\langle\kappa\rangle} > \frac{E_{\rm e}}{\dot{E}_{\rm IC}}.
\eeq
Let us concentrate on the optical soft-photon background with photons at $T\sim4500$~K (i.e., having an average energy $\langle\epsilon\rangle\sim1$~eV). For very energetic leptons, we have to take Klein-Nishina effects into account when calculating the IC loss rate. We thus use the expression of \citet{Ruppel10}
\beq
\dot{E}_{\rm IC} \approx \frac{4\sigma_{\rm T}cu}{3}\frac{\gamma^2_{\rm e}\gamma^2_{\rm KN}}{\gamma^2_{\rm e}+\gamma^2_{\rm KN}},
\eeq
with $\sigma_{\rm T}=6.65\times10^{-25}$~cm$^2$ the Thomson cross section, $u$ the average soft-photon energy density, and 
\beq
\gamma_{\rm KN}\equiv\frac{3\sqrt{5}}{8\pi}\frac{m_{\rm e}c^2}{k_{\rm B}T}
\eeq
the critical Klein-Nishina Lorentz factor. If the particle Lorentz factor $\gamma^2_{\rm e}\gg\gamma^2_{\rm KN}$, the IC loss rate reduces to
\beq
\dot{E}_{\rm IC} \approx \frac{4\sigma_{\rm T}cu\gamma^2_{\rm KN}}{3},
\eeq
yielding
\begin{eqnarray}
\tau_{\rm IC} & \approx & 6\times10^{12}\left(\frac{E_{\rm TeV}T_{4500}^2}{u_{50}}\right)~{\rm s}\nonumber \\
& \approx & 2\times10^5\left(\frac{E_{\rm TeV}T_{4500}^2}{u_{50}}\right)~{\rm yr},
\end{eqnarray}
with $u_{50} \equiv u/(50~{\rm eV/cm^3})$ and $T_{4500}=T/4500~{\rm K}$.
We use $u_{50}$ to scale our results since this value reflects a spatially-averaged value for the energy density; See Figure~\ref{fig:SED} of \citet[][]{BS07} and of \citet[][]{Prinsloo13}.
If we set $R\sim R_{\rm t}\sim10$~pc, we find
\beq
\langle\kappa\rangle < 2.6\times10^{25}\left(\frac{R^2_{10}u_{50}}{E_{\rm TeV}T_{4500}^2}\right)~{\rm cm^2 s^{-1}},\label{eq:1}
\eeq
with $R_{10}\equiv R/10~{\rm pc}$ and $R_{\rm t}$ the tidal radius. This upper limit is similar with the value of the Bohm coefficient at $E_{\rm e}=1$~TeV.
\citet{Kopp13} inferred values for $\kappa$ that are slightly larger at 1~TeV than the Bohm value (for $B_{-6}\sim5$) when fitting the X-ray surface brightness profile, although they assumed an energy dependence $\kappa\propto E_{\rm e}^{0.6}$. They also noted that by assuming Bohm diffusion they could fit the X-ray surface-brightness data, and that the degeneracy in diffusion index and normalization may be broken by using spatial data in a different waveband as well as more spectral data. The caveat is that both the spatial and spectral fit should be reasonable. While \citet{Kopp13} could fit the X-ray surface brightness profile, their predicted SED did not match the data. We thus update their calculation so as to fit both these quantities (Section~\ref{sec:leptonic}).

Additionally, one may argue that since we observe IC emission up to $E_\gamma\sim10$~TeV, we must have
\beq
\tau_{\rm SR} \gtrsim \tau_{\rm IC}.
\eeq
This implies (at those high energies) that
\beq
\dot{E}_{\rm SR} \lesssim\dot{E}_{\rm IC}, 
\eeq
which yields a limit on the magnetic field
\beq
B_{-6} \lesssim 8\left(\frac{\sqrt{u_{50}}}{T_{4500}E_{\rm TeV}}\right).
\eeq

Therefore, from the simple arguments above, we find typical values of $B_{-6}\sim10$ and $\langle\kappa\rangle\sim5\times10^{25}~{\rm cm^2 s^{-1}}$ around $E_{\rm TeV}\sim1$, similar to what was found by \citet{Kopp13}. At these typical cluster $B$-fields, the LESR spectrum should peak around
\beq
E_\gamma = 0.29h\nu_{\rm crit} \approx 2\times10^{-5} B_{-6} E_{\rm TeV}^2~{\rm keV},\label{eq:SR_freq}
\eeq
and for $B_\perp\sim5~\mu$G and $E_{\rm e}\sim 10$~TeV, this component should peak around $\sim0.01~$keV, with $B_\perp=B\sin\alpha^\prime$ and $\alpha^\prime$ the pitch angle. This is consistent with our findings in the next section.

\subsection{Leptonic Modeling of the Broadband SED of \ter}\label{sec:leptonic}
\subsubsection{LESR and IC Components}
\begin{figure*}
		\includegraphics[width=0.85\textwidth]{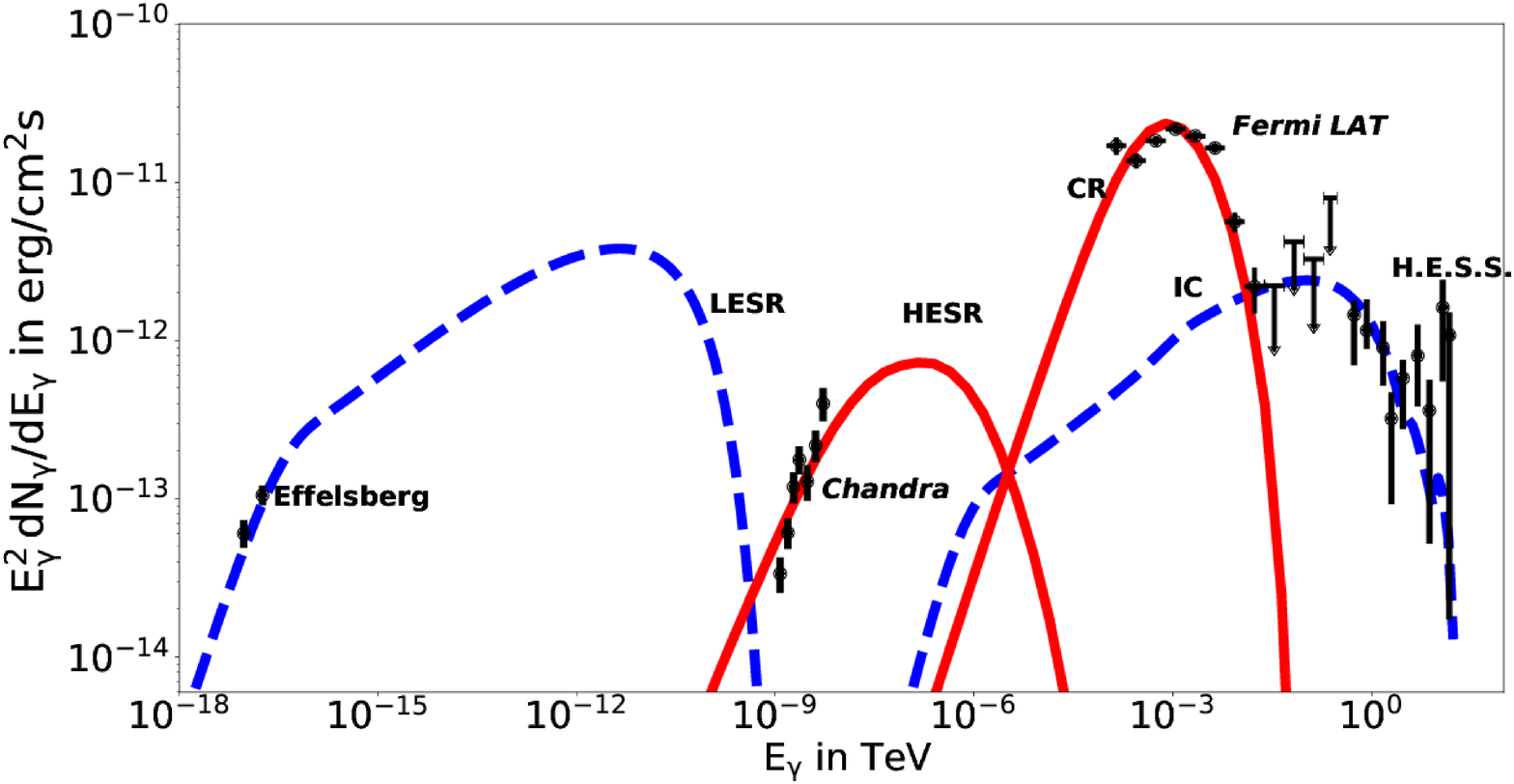}
		\centering
	\caption{Different spectral components for \ter predicted by the leptonic models of \citet{Kopp13} and \citet{Harding08,HK15}. Using the first model, we calculate the low-energy SR (LESR) and VHE IC components (integrated over all $r_{\rm s}$; dashed blue lines). We assumed $E_{\rm e,min} = 9\times10^{-3}$~TeV, $E_{\rm e,max} = 10$~TeV, $Q_{0} = 1.4\times10^{34}~{\rm erg}^{-1}~{\rm s}^{-1}$, $B = 4.0\,\mu \rm G$, $\Gamma = 1.5$, and $\kappa = 7\times10^{-5}\, \rm kpc^{2}\,Myr^{-1}\approx2\times10^{25}~{\rm cm}^2~{\rm s}^{-1}$. We used a distance of $d=5.9\,{\rm kpc}$, core radius $R_{\rm c}=0.15^\prime=0.26\,{\rm pc}$, half-mass radius $R_{\rm hm}=0.52^\prime=0.89\,{\rm pc}$, and tidal radius $R_{\rm t}=4.6^\prime=7.9\,{\rm pc}$. The HESR and CR components (red lines) are predictions using the model of \citet{Harding08,HK15} for $\langle\alpha\rangle = 45^\circ$, $\langle\zeta\rangle = 60^\circ$, $\langle P \rangle = 7.7\times 10^{-3}$~s, and $\langle B_{\rm s}\rangle=5.8\times10^9$~G. We also indicate \textit{Chandra} \citep{Eger10}, H.E.S.S. \cite{Abramowski11}, and radio data \citep[``Region~11'' as defined by][]{Clapson2011}. The uncertainties in our LAT points do not reflect possible systematic errors on the Galactic diffuse emission model.}
	\label{fig:SED}
\end{figure*}

We present new spectral fits\footnote{The main aim of this paper is to ascertain whether we can elucidate the broadband spectral emission properties of Terzan 5 as well as those of the underlying sources that inject particles into Terzan 5. However, we realize that the energy-dependent morphology of this cluster is quite complex, so much so that it challenges the idea of a single (collective) particle population injected by the MSPs being responsible for all spectral emission components originating from partially-overlapping spatial regions of different extents. Yet, to facilitate usable conclusions to be drawn from the current data, we do invoke a single population and study the source energetics, while deferring a study of spatial properties of Terzan 5 to future work.} to the SED of \ter using the model of \citet{Kopp13} as shown in Figure~\ref{fig:SED} (blue dashed lines). The model includes a spatial dimension, refined stellar soft-photon energy density profile and full particle transport, taking diffusion and radiation losses into account with the assumptions of spherical symmetry and a steady-state regime.

In Figure~\ref{fig:SED} we indicate radio data (labeled ``Effelsberg'') associated with ``Region~11'' (a prominent, large-scale, asymmetric feature offset from the center) as defined by \citet{Clapson2011}. We fit these points with our predicted LESR component, in keeping with the suggestion by \citet{Clapson2011} that the flux from this region may be due to unpulsed SR from leptons that were injected by the MSPs into the GC and diffused throughout the cluster.
Our predicted LESR component is much below the estimated BB flux level in the optical band (see Section~\ref{optical}). As mentioned in Eq.~(\ref{eq:SR_freq}), we expect this component to peak around 1~keV for particle energies $E_{\rm e}\sim100$~TeV and $B\sim10\,\mu$G. This led \citet{Kopp13} to fit the X-ray surface brightness profile measured by \textit{Chandra} in order to constrain the diffusion coefficient to $\kappa\sim 3.3\times10^{25}$~cm$^2$\,s$^{-1}$ at 1~TeV, similar to Eq.~(\ref{eq:Bohm}) and Eq.~(\ref{eq:1}). However, although their model prediction reproduced the flux level at a few keV, they could not fit the spectral slope of the observed data. This implies that the observed diffuse X-ray emission may be due to a different spectral component. We therefore now choose the maximum particle energy $E_{\rm e,max}=10$~TeV and a slightly lower $B$-field of $B=4\,\mu$G so that our new LESR component peaks around $E_\gamma\sim0.01$~keV (as we found in Section~\ref{sec:model}) and thus cuts off below the \textit{Chandra} data. Furthermore, the low-energy tail of our predicted IC component satisfies the new \textit{Fermi} data and upper limits and we also reproduce the H.E.S.S.\ data. 

\subsubsection{Primary CR and Pair High-energy SR Components}
We use the model of \citet{HK15} to fit the GeV and keV data. Similar to previous studies \citep[e.g.,][]{HUM05,Venter09_GC,Zajczyk13}, we fit the \textit{Fermi} LAT data using the cumulative primary CR component of pulsed $\gamma$-ray emission originating in the MSP magnetospheres (GeV component indicated by a solid red line and labeled ``CR'' in Figure~\ref{fig:SED}). This has been a standard interpretation for the GeV spectrum measured by the LAT for several GCs \citep{Abdo10}.

Following the idea of \citet{Kopp13}, we propose that the \textit{Chandra} data indicate the presence of a ``new'' high-energy SR (HESR) component that has not been modeled in detail in this context\footnote{As a practical measure, we attribute the \textit{Chandra} data solely to collective, pulsed, non-thermal, magnetospheric pulsar emission. There could be contributions by other sources, but we do not know \textit{a priori} the properties of unresolved stellar members hosted by Terzan 5.} before: the cumulative pulsed SR from pairs generated within the magnetospheres of the host MSPs, radiating at altitude that are a substential fraction of the light cylinder (at a radius $R_{\rm LC}=c/\Omega$, with $\Omega$ the angular speed, where the co-rotational speed equals $c$) through cyclotron resonant absorption of radio photons \citep[cf.][]{Harding08}. Given the much higher local $B$-field (e.g., the magnetospheric field at the MSPs' light cylinder may reach $B_{\rm HESR}\sim10^6$~G for the most energetic ones\footnote{The average $B$-field at the light cylinder may be closer to $\sim10^4$~G; however, the pair SR component is likely dominated by the MSPs with the highest spin-down luminosities and $B$-fields.} vs.\ the much lower GC field $B_{\rm LESR}\sim10^{-5}$~G) and the much smaller average pitch angle ($\alpha_{\rm HESR} \sim0.1$ vs.\ $\alpha_{\rm LESR}\sim\pi/2$ radians) as well as different particle energies, the cutoff energy of this new component is much higher than that of the LESR spectrum:
\beqa
\frac{E_{\rm HESR,cut}}{E_{\rm LESR,cut}} & \sim & \left(\frac{\gamma_{\rm SR}}{\gamma_{\rm LESR}}\right)^2\frac{B_{\rm HESR}\sin\alpha_{\rm HESR}}{B_{\rm LESR}\sin\alpha_{\rm LESR}}\\
& \sim & \left(\frac{10^4}{10^7}\right)^2\frac{10^6~{\rm G}\times10^{-1}}{10^{-5}~{\rm G}}\sim 10^4.
\eeqa
This simple scaling predicts a cutoff $E_{\rm HESR,cut}\lesssim100$~keV, and thus provides us with a low-energy tail that might fit the X-ray data. This idea is also supported by observations of sources embedded in 47~Tucanae: \citet{Bogdanov08} noted that even though most of the observed MSPs exhibit soft thermal spectra, three of them manifest hard power-law components. These components may plausibly be attributed to binary shock emission or magnetospheric SR. It is furthermore supported by detection of hard non-thermal X-ray emission from a number of field MSPs.

As a proof of principle, we now calculate model spectra invoking a cumulative pulsed HESR component originating in the MSP magnetospheres to fit the \textit{Chandra} data (keV component indicated by a solid red line and labeled as ``HESR'' in Figure~\ref{fig:SED}). We use a force-free $B$-field in the inertial observer frame, choosing a slot gap width of 0.03$\Theta_{\rm PC}$, with $\Theta_{\rm PC}$ the polar cap angle (the inner and outer angular boundaries of the gap were set at open-volume coordinates $r_{\rm ovc}\in(0.90,0.93)$, where the $r_{\rm ovc}$ coordinate labels self-similar rings, $r_{\rm ovc}=0$ being the magnetic pole and $r_{\rm ovc}=1$ being the polar cap rim; see \citealt{Dyks04}), and a constant $E$-field from the MSP surface to $2R_{\rm LC}$, set by an inverse acceleration length scale of $R_{\rm acc}=d\gamma_{\rm e}/dl=2$ cm$^{-1}$ (i.e., $E_{||}=R_{\rm acc}m_{\rm e}c^2/e$, with $\gamma_{\rm e}$ the particle Lorentz factor and $dl$ the step length along the particle trajectory, $m_{\rm e}$ and $e$ the electron mass and charge). We divide the fraction of stellar surface covered by $B$-field line footpoints that are within the gap using 4 self-similar rings and 72 azimuthal divisions. We choose an average pulsar period of $\langle P\rangle = 7.7$~ms, and by fixing the average surface $B$-field to $\langle B_{\rm s}\rangle=5.8\times10^9$~G and moment of inertia $\langle I \rangle=1.56\times10^{45}$~g\,cm$^2$, we obtain $\langle \dot{P}\rangle \sim 7\times10^{-19}$\,s\,s$^{-1}$ and $\langle\dot{E}\rangle\sim9.08\times10^{34}$~erg\,s$^{-1}$. The latter may represent significant contributions from more energetic MSPs. Nonetheless, we take these values as representative\footnote{Unfortunately, there is a large  uncertainty in the MSP population's properties. While a full Monte Carlo investigation of the SED may be preferable from a first-principles point of view, this will introduce many more uncertainties and a large range for the SED components' shapes and levels, so this will probably not lead to any conclusive answers. We therefore deem the approach of studying the behaviour of an ``average MSP'' as the most practical, although we are cognisant of the fact that a particularly powerful MSP may skew the results.} of the pulsars in \terp We furthermore use a polar cap pair spectrum calculated for an offset-polar-cap $B$-field \citep{Harding11a,Harding11b,Barnard16} with an offset parameter of $\epsilon = 0.6$. We choose an average magnetic inclination angle of $\langle\alpha\rangle=45$\dgr and average observer angle of $\langle\zeta\rangle=60$\dgr (for both HESR and CR components). See \citet{HK15} for details. 

The number of visible $\gamma$-ray pulsars ($N^{\rm \gamma}_{\rm vis}$) is constrained by the primary CR flux level, for a given set of model parameters. Alternatively, if we fix $N^{\rm \gamma}_{\rm vis}=N^{\rm rad}_{\rm vis}=37$ to the number of visible radio pulsars (since nearly all currently detected $\gamma$-ray MSPs by \textit{Fermi} are radio-loud), we may constrain other parameters such as the gap width and average pulsar geometry $\alpha$ and $\zeta$, or $\langle P\rangle$ and $\langle\dot{P}\rangle$. Unfortunately, it is difficult to break this degeneracy using X-ray data, since one may expect that $N^{\rm X}_{\rm vis}\lesssim N^\gamma_{\rm vis}$ if their X-ray beams are slightly narrower than the $\gamma$-ray ones, and equality may not hold exactly. One may additionally write that $N^{\rm X}_{\rm tot} = N^{\rm X}_{\rm vis}+N^{\rm X}_{\rm invis}\geq N^{\rm rad}_{\rm vis}=37$ and $N^{\rm \gamma}_{\rm vis}\geq 37$.
The product $M_\pm N^{\rm X}_{\rm vis}$ is being set by the LESR flux level, so these two parameters are degenerate. Using the HESR (\textit{Chandra}) flux level, we constrain the product $N^{\rm X}_{\rm vis}\langle M_\pm\rangle\sim1.9\times10^4$, with $\langle M_\pm\rangle$ the average number of pairs produced per primary extracted from the polar cap, per pulsar (the average electron-positron pair multiplicity). If we take  $N^{\rm X}_{\rm vis}\approx 35$, we obtain $\langle M_\pm\rangle\approx540$. However, this value depends crucially on the assumptions of the magnetospheric model: more optimistic assumptions about the electrodynamics (e.g, a higher $B$-field or current, that will influence the particle transport) may lead to a larger single-MSP spectrum, and yield a lower constant (value for the product of $N^{\rm X}_{\rm vis}\langle M_\pm\rangle$), thus lowering the value for $\langle M_\pm\rangle$.

Previously, \citet{Kopp13} found an optimal source strength of $Q_0 \sim 6\times10^{33}$~erg$^{-1}$s$^{-1}$ when fitting the LESR and IC components. The value of $Q_0$ is usually constrained by assuming a parametric form for the particle injection spectrum
\begin{equation}
  Q(E_{\rm e}) = Q_0E_{\rm e}^{-\Gamma}
\end{equation}
and using conservation of charge and energy per unit time (i.e., conservation of current and luminosity; \citealt{Buesching08,Venter15c}):
\begin{eqnarray}
	\int_{E_{\rm e,min}}^{E_{\rm e,max}}Q(E_{\rm e})\,dE_{\rm e} & = & N_{\rm MSP,tot}\times\nonumber\\
	& & \left(\langle M_\pm\rangle + 1\right)\langle\dot{n}_{\rm GJ}\rangle\\
	\int_{E_{\rm e,min}}^{E_{\rm e,max}}E_{\rm e}Q(E_{\rm e})\,dE_{\rm e} & = & N_{\rm MSP,tot}\eta_{\rm p}\langle\dot{E}\rangle,
\end{eqnarray}
with $\langle\dot{n}_{\rm GJ}\rangle=4\pi^2B_{\rm s}R^3/ceP^2\propto \langle\dot{E}\rangle^{1/2}$ the average Goldreich-Julian rate of particles injected per second for a pulsar period $P$, surface magnetic field $B_{\rm s}$ and stellar radius $R$ \citep{GJ69}, and $\eta_{\rm p}$ the efficiency of converting the average spin-down luminosity to particle power. The ``+1'' in the first equation above represents the contribution from primary particles. The above system of equations may have up to 10 free parameters, implying a large degeneracy of parameters. We found an optimal value of $Q_0 \sim 1.4\times10^{34}$~erg$^{-1}$s$^{-1}$ (Fig.~\ref{fig:SED}) by fitting the unpulsed spectral components (for particular choices of other free parameters, e.g., $\kappa$ and $B$, and using $\langle\dot{E}\rangle=9.08\times10^{34}$~erg\,s$^{-1}$ and $\eta_{\rm p}=3\%$ and $N_{\rm MSP,tot}\sim40$). This leads to a constraint on the average multiplicity:
\begin{eqnarray}
	\langle M_\pm\rangle & = & \frac{Q_0\left(E_{\rm e,max}^{1-\Gamma}-E_{\rm e,min}^{1-\Gamma}\right)}{(1-\Gamma)N_{\rm MSP,tot}\langle\dot{n}_{\rm GJ}\rangle} - 1\nonumber\\
	& \approx & 20\left(\frac{\eta_{\rm p}}{3\%}\right)\!\!\!\left(\frac{2.7\times10^{32}~{\rm s}^{-1}}{\langle\dot{n}_{\rm GJ}\rangle}\right)\left(\frac{\langle\dot{E}\rangle}{9\times10^{34}~\rm erg\,s^{-1}}\right)\\
       & \propto & \langle\dot{E}\rangle^{1/2},\!\!\!\!\!\!\!\!\!\label{eq:M1} \nonumber
\end{eqnarray}
with the value of $\langle\dot{n}_{\rm GJ}\rangle$ reflecting the choice for the average $\langle P \rangle$ and $\langle B_{\rm s}\rangle$ of the MSPs as mentioned earlier, for consistency. This estimate of $\langle M_\pm\rangle\approx20$ is quite a bit lower than the previous one of $\langle M_\pm\rangle\approx540$ as inferred from the HESR component. There are ways to mitigate this difference, given the uncertainty and degeneracy in several model parameters. The estimate of $\langle M_\pm\rangle$ using the unpulsed spectral components may be raised to $\langle M_\pm\rangle\approx60$ by using $E_{\rm min}\sim40$ GeV, $E_{\rm max}\sim7$ TeV, and $\Gamma = 1.6$, without significantly changing the SED. Next, the discrepancy can be lowered to a factor $\sim4.5$ by increasing $\langle I \rangle$ by a factor of $\sim2$, since $\langle M_\pm\rangle \propto \dot{E}^{1/2}$ for the unpulsed case, while $\langle M_\pm\rangle \propto \langle\dot{E}\rangle^{-1/2}$ for the pulsed case. This, however, raises $Q_0 \propto \langle\dot{E}\rangle$ by a factor of $\sim2$ so that LESR overshoots the data slightly, but this effect can then be mitigated by choosing $B \approx1~\mu$G. Lastly, the remaining discrepancy of a factor of $\sim4.5$ can be ameliorated by assuming a larger value for $\epsilon \sim~0.7$ (implying more pairs) and a larger gap width (say, increasing the upper boundary to $r_{\rm ovc}\sim0.96$, implying a larger active area on the stellar surface and thus lowering the demand on $\langle M_\pm\rangle$). There are also uncertainties in the angles $\langle\alpha\rangle$ and $\langle\zeta\rangle$ that may have a significant effect on the HESR flux. Lastly, using average values $\langle B_{\rm s}\rangle$ and $\langle P\rangle$ leads to average values for $\langle \dot{n}_{\rm GJ}\rangle$ and $\langle\dot{E}\rangle$, and this introduces further uncertainty. It is thus possible to pick (non-unique combinations of) values for some model parameters that would make the two estimates of $\langle M_\pm\rangle$ (using the pulsed and unpulsed SED components) consistent with each other, without violating the observed SED.
The actual value of $M_\pm$ for MSPs is quite uncertain. Polar cap pair cascades in a pure dipole field give very low values of $M_\pm$ for the bulk of MSPs, which prompted the suggestion that distortions of the B-field near the neutron star could increase $M_\pm$ \citep{Harding11b} but the magnitude and structure of such distortions are not known. Comparing results of particle-in-cell simulations \citep{Kala18} with $\gamma$-ray spectral cutoffs seen in \textit{Fermi} pulsars can give estimates of MSP $M_\pm$ needed to screen the global electric fields. This study indicates that the estimated MSP $M_\pm$ span a large range from $1 - 10^3$.
   
\subsubsection{Balancing the Energetics of the MSP Population}
Our model provides reasonable fits to the \textit{Chandra} and \textit{Fermi} data for typical model parameters. However, one also has to consider whether this scenario is plausible in terms of energetics and the sensitivity of \textit{Chandra}, i.e., would \textit{Chandra} have seen these ``unresolved MSPs'' postulated by the model to explain the diffuse X-ray flux seen by \citet{Eger10}, or can one indeed explain the observed SR flux by a reasonable number of visible and invisible (unresolved) MSPs? The answer to this question lies in the (uncertain) population properties and emission energetics of the MSPs. We investigate this question by taking two approaches below.

From the \textit{Chandra} data analysis, we can obtain three constraints. \citet{Eger10} assume a point-source sensitivity of $\sim2\times10^{-15}$~erg\,s$^{-1}$\,cm$^{-2}$ in the 0.5 $-$ 7.0~keV band. This leads to the first constraint of the minimum detectable luminosity of (i) $L_{\rm X, Chandra}\sim 7\times10^{30}$~erg\,s$^{-1}$ for their assumed distance of $d=5.5$~kpc. This is similar to the value of $L_{\rm X,Chandra}\sim (1-3)\times10^{31}$~erg\,s$^{-1}$ for an assumed distance of $d=8.7$~kpc found by \citet{Heinke06}. Let us adopt the first value. Second, \citet{Eger10} note that the total observed unabsorbed diffuse excess luminosity\footnote{We note that the power-law fit to the data implies that the visible non-thermal luminosity is $L_{\rm X,vis}=8.52\times 10^{32}$~erg\,s$^{-1}$ \citep{Eger10}, using data from annuli lying between $55^{\prime\prime}$ and $174^{\prime\prime}$.
By integrating our predicted $E_\gamma dN/dE\gamma$ HESR spectrum in the 1 $-$ 7~keV band, we find $L_{\rm X,HESR}\sim5.5\times10^{32}$~erg\,s$^{-1}$. This is close to this luminosity, with the discrepancy explained by the fact that the model does not perfectly match the data in terms of the spectral slope. However, this power-law-luminosity is a factor $\sim2$ lower than the total observed luminosity as noted in the main text, which is also the number quoted by \citet{Eger10} in their interpretation section. We decided to use the higher value, following this usage by \citet{Eger10}, and note that if we use the lower value, the solutions in Table~\ref{tab:pop} have similar best-fit parameters but with lower MSP numbers reflecting the lower value of $L_{\rm X,vis}$ in this case.} is $L_{\rm X,tot}=2\times10^{33}$~erg\,s$^{-1}$, and estimate that the contribution of unresolved point sources\footnote{We use the label ``invisible'' in what follows to refer to those pulsars that have too low a spin-down luminosity to be detectable as single point sources by \textit{Chandra}, but that may contribute to the cumulative unresolved point-source luminosity as a population of low-energetic pulsars. As before, we discard the contribution of other source classes to this unresolved luminosity.} in the $1^\prime - 3^\prime$ region is $=7\times10^{31}$~erg\,s$^{-1}$. We thus set (ii) $L_{\rm X,vis}=2\times10^{33}$~erg\,s$^{-1}-7\times10^{31}$~erg\,s$^{-1}=1.93\times10^{33}$~erg\,s$^{-1}$ and (iii) $L^{\rm X}_{\rm invis}=7\times10^{31}$~erg\,s$^{-1}$. In order to convert X-ray luminosities to pulsar spin-down values, one needs an efficiency factor $\eta_{\rm X}$:
\begin{eqnarray}
    L_{\rm X,vis} & = & \eta^{\rm  X}_{\rm vis}N^{\rm X}_{\rm vis}\langle \dot{E} \rangle_{\rm vis}\\
    L_{\rm X,invis} & = & \eta^{\rm  X}_{\rm invis}N^{\rm X}_{\rm invis}\langle \dot{E}\rangle_{\rm invis},\label{eq:balance}
\end{eqnarray}
with the total number of MSPs $N^{\rm X}_{\rm tot}=N^{\rm X}_{\rm vis} + N^{\rm X}_{\rm invis}\geq N^{\rm rad}_{\rm vis}=37$. This is a very unconstrained system of equations. To simplify this, one may assume that the whole population of MSPs may be characterized by a single $\eta_{\rm  X}=\eta^{\rm  X}_{\rm vis}=\eta^{\rm  X}_{\rm invis}$. Division of the former equation by the latter and fixing $\langle \dot{E} \rangle_{\rm vis}$ then yields the following constraint:
\begin{equation}
N^{\rm X}_{\rm vis}\langle \dot{E} \rangle_{\rm vis} = k_{\rm L}N^{\rm X}_{\rm invis}\langle \dot{E}\rangle_{\rm invis},
\end{equation}
with $k_{\rm L} = L_{\rm X,vis}/L_{\rm X,invis}$ being a constant. While there is some degeneracy, this constraint may, e.g., be satisfied for the following choices: 
$\langle \dot{E} \rangle_{\rm vis}=9.08\times10^{34}$~erg\,s$^{-1}$, $\langle \dot{E} \rangle_{\rm invis}=8\times10^{33}$~erg\,s$^{-1}$, $N^{\rm X}_{\rm vis}=41$ and $N^{\rm X}_{\rm invis}=17$ (implying $\eta_{\rm  X}=0.05\%$). If we adopt $\langle \dot{E} \rangle_{\rm vis}=1.8\times10^{34}$~erg\,s$^{-1}$ \citep[e.g.,][]{Abdo10}, the following values satisfy the constraint above: $\langle \dot{E} \rangle_{\rm invis}=10^{33}$~erg\,s$^{-1}$, $N^{\rm X}_{\rm vis}=23$ and $N^{\rm X}_{\rm invis}=15$ (implying $\eta_{\rm  X}=0.5\%$); alternatively, we can set $\langle \dot{E} \rangle_{\rm vis}=10^{34}$~erg\,s$^{-1}$, obtaining $\langle \dot{E} \rangle_{\rm invis}=2.8\times10^{32}$~erg\,s$^{-1}$, $N^{\rm X}_{\rm vis}=19$ and $N^{\rm X}_{\rm invis}=25$ (implying $\eta_{\rm  X}=1\%$). These numbers seem reasonable and adhere to the constraint that $N^{\rm X}_{\rm tot} = N^{\rm X}_{\rm vis} + N^{\rm X}_{\rm invis}\geq N^{\rm rad}_{\rm vis}=37$. Thus, we consider our assumption of attributing the X-ray emission to the cumulative pair SR, as used in the previous section, plausible.

In an attempt to perform a more robust analysis, potentially obtain stronger constraints on the MSP population and break some parameter degeneracies, we consider a parametrized pulsar spin-down luminosity function $N_{\rm MSP}(>\!\!\dot{E})\propto \dot{E}^{-\gamma_{\rm L}}$ \citep{Johnston1996}. We will use this to balance the required X-ray energetics by assuming that $\dot{E}_{\rm vis}\propto L_{\rm X,vis}$ and $\dot{E}_{\rm invis}\propto L_{\rm X,invis}$. This implies $dN/d\dot{E}=N^\prime_0(\dot{E}/\dot{E}_0)^{-(\gamma_{\rm L} + 1)}$, with $N^\prime_0$ a normalization constant. \citet{Johnston1996} infer a typical GC value of $\gamma_{\rm L}\sim0.5$, while \citet{Heinke06} find $\gamma_{\rm L}\sim0.4 - 0.7$ for \terc depending on the energy band. By defining $\dot{E}_{\rm b} = L_{\rm X,Chandra}/\eta_{\rm X}$, one can next recover the following quantities:
\begin{eqnarray}
N^{\rm X}_{\rm tot} & = &\int_{\dot{E}_{\rm min}}^{\dot{E}_{\rm max}}\left(\frac{dN}{d\dot{E}}\right)\,d\dot{E}, \\
N^{\rm X}_{\rm vis} & = &\int_{\dot{E}_{\rm b}}^{\dot{E}_{\rm max}}\left(\frac{dN}{d\dot{E}}\right)\,d\dot{E}, \\
N^{\rm X}_{\rm invis} & = & \int_{\dot{E}_{\rm min}}^{\dot{E}_{\rm b}}\left(\frac{dN}{d\dot{E}}\right)\,d\dot{E},  \\
\langle\dot{E}\rangle_{\rm vis} & = & \frac{1}{N^{\rm X}_{\rm vis}}\int_{\dot{E}_{\rm b}}^{\dot{E}_{\rm max}}\dot{E}\left(\frac{dN}{d\dot{E}}\right)\,d\dot{E},\label{eq:Edotvis}\\
\langle\dot{E} \rangle_{\rm invis} & = & \frac{1}{N^{\rm X}_{\rm invis}}\int_{\dot{E}_{\rm min}}^{\dot{E}_{\rm b}}\dot{E}\left(\frac{dN}{d\dot{E}}\right)\,d\dot{E}.
\end{eqnarray}
We want to solve for four quantities: $\dot{E}_{\rm min}$,
$\dot{E}_{\rm max}$, $N^\prime_0$ (or equivalently $N^{\rm X}_{\rm tot}$), and $\gamma_{\rm L}$; once these are fixed, we can infer the MSP population properties through the above equations. We note, however, that we are using this luminosity function to fit X-ray luminosities, which are integral quantities. We therefore expect to find degenerate solutions as different combinations might yield the same integral luminosities. Thus, we need four constraints or measurements. We can use the same three constraints as before. Crucially, one needs to specify a fourth parameter $\eta_{\rm X}$ to convert from spin-down luminosities to X-ray luminosities. By fixing $\eta_{\rm X}$, we implicitly fix the product $N^{\rm X}_{\rm vis}\langle\dot{E}\rangle_{\rm vis}$. As a first attempt, let us assume $\eta_{\rm X}=0.05\%$ (e.g., for $N^{\rm X}_{\rm vis}=41$ and $\langle\dot{E}\rangle_{\rm vis}=9.08\times10^{34}$~erg\,s$^{-1}$ to make the calculation consistent with the previous estimate). It is difficult to obtain the actual value for $\langle\dot{E}\rangle_{\rm vis}$, given the effect of the GC cluster potential on the $\dot{P}$ of each MSP \citep[e.g.,][]{Bogdanov08}. If we could, this would further constrain the system via Eq.~(\ref{eq:Edotvis}). \citet{Heinke06} note that while they did not detect an X-ray MSP explicitly, one X-ray source could plausibly be an MSP based on the proximity to a radio MSP position; they also noted that more identifications of X-ray MSPs could be made as radio positions become available. We thus have additional constraints $N^{\rm X}_{\rm vis}\gtrsim 1$ and $N^{\rm X}_{\rm tot}\geq N^{\rm rad}_{\rm vis}$ (which may be used as checks on the consistency of the solutions we obtain). We do obtain a non-unique solution for each fixed value of $\eta_{\rm X}$. However, $\eta_{\rm X}$ is not known and the parameters that satisfy the other three constraints are quite degenerate, as expected. Table~\ref{tab:pop} indicates a number of parameter combinations that satisfy the observational constraints\footnote{A preliminary Markov-chain Monte Carlo investigation \citep{EMCEE} confirmed the degenerate nature of the free parameters (some are correlated) as well as them being quite unconstrained (reflected by asymmetrical and flat probability distributions, as well as elongated confidence contours). Best-fit values furthermore depend on the choice of priors / parameter bounds. The median values are, however, similar to those in Table~\ref{tab:pop}.}. It is clear that a different choice of $\eta_{\rm X}$ will favor a different solution that will imply a different
value of $\langle\dot{E} \rangle_{\rm vis}$ (which also depends on the the average moment of inertia $\langle I\rangle$, $\langle P\rangle$, and $\langle \dot{P}\rangle$). For example, a higher value of $\eta_{\rm X}$ will yield a lower value of $\langle\dot{E} \rangle_{\rm vis}$ or $\langle I\rangle$ for a given value of $N^{\rm X}_{\rm vis}$ and keeping other parameters fixed. If we require $\langle\dot{E} \rangle_{\rm vis}$ to be the same as assumed in the model used to predict the pulsed emission, this may lead to unrealistic values for $\gamma_{\rm L}$, for a given $\eta_{\rm X}$. Relaxing this requirement (which may easily be done, given other parameter uncertainties) implies more suitable values for the other parameters. It is therefore clear that the system of equations is very coupled and the parameters are degenerate, given the lack of suitable constraints. One may think to constrain the solution space by requiring $N^{\rm X}_{\rm tot}=N^{\gamma}_{\rm tot}\approx N^{\gamma}_{\rm vis}=180^{+120}_{-100}$, the latter being the estimated total number of visible MSPs in \ter as inferred from the \textit{Fermi}-measured GeV energy flux \citep{Abdo10}. However, this estimate is quite uncertain and does not contain uncertainties in distance (the square of which determines the $\gamma$-ray luminosity $L_\gamma$) and conversion efficiency of $\dot{E}_{\rm vis}$ to $L_\gamma$, so this does not seem to be a strong constraint. Likewise, we chose $N^{\gamma}_{\rm tot}=37$ when fitting the CR component, but this value is also subject to other model assumptions such as MSP geometry and gap width. Finally, it seems that the last few entries in Table~\ref{tab:pop} might be the more plausible combinations in view of the independent constraints on $\gamma_{\rm L}\sim0.4 - 0.7$ and $N^{\rm X}_{\rm vis}\gtrsim 1$ (i.e., probably relatively small numbers of visible X-ray pulsars) gleaned from the analysis of \citet{Heinke06}. Thus, the uncertainty in several parameters, particularly $\eta_{\rm X}$, as well as parameter degeneracies preclude us from making definite statements about the MSP population properties. Yet, we see that there are several plausible solutions that characterize and constrain the MSP population's energetics, implying that the scenario of MSPs being responsible for the broadband SED may be justified and thus be plausible.

\citet{Gonthier2018} derive a luminosity function for MSPs in 47~Tucanae through a population synthesis that fits the \textit{Fermi} spectrum, presumed to be the combined emission from all MSPs in the cluster.  They find a $\gamma$-ray luminosity distribution that peaks at $\sim 10^{33}$\,erg\,s$^{-1}$, spin-down power distribution peaking at $\sim 2 \times10^{34}$\,erg\,s$^{-1}$, extending down to $\sim 10^{30}$\,erg\,s$^{-1}$, with a $\gamma$-ray efficiency around 0.1. Their peak $\dot{E}$ value is similar to the $\langle\dot{E}_{\rm vis}\rangle$ we have derived for \terp

\begin{deluxetable*}{cccccccccccc}
\tablecaption{Sample parameter combinations that lead to a balance of the X-ray-implied energetics.\label{tab:pop}}
\tablecolumns{9}
\tablehead{\colhead{$\eta_{\rm X}$} & \colhead{$\langle\dot{E}\rangle_{\rm invis}$} & \colhead{$\langle\dot{E}\rangle_{\rm vis}$} &\colhead{$\dot{E}_{\rm min}$} &\colhead{$\dot{E}_{\rm max}$} &\colhead{$\gamma_{\rm L}$} &\colhead{$N^{\rm X}_{\rm vis}$} &\colhead{$N^{\rm X}_{\rm invis}$} &\colhead{$N^{\rm X}_{\rm tot}$}}
\startdata
0.05\% & $3.0\times 10^{33}$ & $9.2\times 10^{34}$ & $10^{31}$ & $2.4\times 10^{35}$ & $-0.19$ & 43 & 45 & 88\\
0.05\% & $3.5\times 10^{32}$ & $1.8\times 10^{35}$ & $10^{29}$ & $10^{36}$ & $0.21$ & 22 & 399 & 421\\
0.5\% & $1.2\times 10^{32}$ & $3.8\times 10^{34}$ & $10^{31}$ & $10^{36}$ & $0.50$ & 10 & 116 & 126\\
0.5\% & $1.5\times 10^{32}$& $2.9\times 10^{34}$ & $10^{31}$ & $3.6\times 10^{35}$ & $0.40$ & 14 & 96 & 110\\
1\% & $7.4\times 10^{31}$ & $2.5\times 10^{34}$ & $10^{31}$ & $2.0\times 10^{36}$ & $0.60$ & 8 & 95 & 103\\
1\% & $1.3\times 10^{32}$ & $2.4\times 10^{34}$ & $3.0\times 10^{31}$ & $2.9\times 10^{36}$ & $0.64$ & 8 & 53 & 61\\
1\% & $2.5\times 10^{32}$ & $2.0\times 10^{34}$ & $10^{32}$ & $3.0\times 10^{36}$ & $0.69$ & 10 & 27 & 37
\enddata
\tablecomments{The units of the spin-down luminosities are erg\,s$^{-1}$.}
\end{deluxetable*}
 
The bulk of visible radio MSPs occur in the core of the cluster, and one expects the majority of pulsars here due to the deep potential well of the GC. However, one may expect to find a small number of MSPs farther out, depending on their birth and evolutionary history. The fact that the diffuse X-ray flux profile measured by \citet{Eger10} drops off slightly slower than the generalized King profile fit \citep[e.g.,][]{King62} to the detected X-ray point source distribution \citep{Heinke06} as well as the infrared surface brightness profile measured by \citet{Trager1995} may support this idea, and a (slowly) decreasing MSP density with radius may plausibly correlate with the observed decreasing X-ray flux profile. \citet{Eger10} detected non-thermal X-ray emission beyond the half-mass radius of \terp If we take the above energetics argument as plausible, this would imply possibly tens of unresolved MSPs and a handful MSPs that are in principle visible in X-rays in this region. This would also imply that even more MSPs should be visible in X-rays at the core vs.\ outer reaches of the GC, but we do not have constraints on the diffuse X-ray emission at the GC centre at this stage, and source confusion in this dense region may complicate the matter. Future constraints on the central diffuse X-ray emission, the spatial distribution of the MSP population, and the average expected multiplicity and spin-down power will thus more deeply probe our hypothesis that the HESR component is due to magnetospheric, pulsed SR from pairs.

\section{CONCLUSIONS}
\label{sec:concl}
The main focus of this paper has been two-fold: to gather more data on \ter and to scrutinize ideas about the particle sources and emission processes responsible for the broadband emission spectrum we observe from this cluster. Our models postulated four spectral components (LESR, HESR, CR and IC) and attempted to constrain the MSP population's distribution of spin-down luminosity using the observed X-ray diffuse emission. 

We obtained new \textit{Fermi} data that we could fit using a model for the cumulative CR from a population of MSPs embedded within \terp  These data also proved to be constraining for the low-energy tail of the unpulsed IC component, yielding a particle efficiency of $\eta_{\rm p}\sim3\%$, depending on the choice of several parameters, notably $\langle\dot{E}_{\rm vis}\rangle$ and $N_{\rm MSP,tot}$.

We demonstrated that we could fit the radio spectral points by invoking an LESR component that might extend into the optical range. We furthermore argued that our predicted LESR flux is far below the expected thermal optical flux level. Thus, obtaining an upper limit on the non-thermal flux in the optical band would be very difficult, given the roughly $N_*\sim10^{5-6}$ point sources that have to be subtracted from an optical map of the GC. Even when performing and subtracting a King model fit to the surface brightness profile, the uncertainty on the remaining diffuse flux would be very large. However, since the BB spectrum occurs over a much narrower energy range than the LESR, there may yet be hope of detecting the latter outside of the optical range.

\citet{Bednarek16} concur with our prior predictions \citep{Venter08,Venter09_GC,Kopp13} that GCs should typically have SR components that peak in the optical / ultraviolet range, but also points out the problem of the dominating radiation field produced by the large population of GC stars. They furthermore mention that quite atypical parameters (a combination of very large cluster $B$-fields and particle energies) would be needed in order to produce an observable level of X-ray flux. Lastly, it would be problematic to compare optical and X-ray brightness profiles, since the underlying source distributions have different spatial and emission characteristics. The respective telescope point spread functions also differ, compounding the problem. Also, the observed \textit{Chandra} spectrum is not well fit by a single LESR component.
To solve these problems and still fit the observed data, we invoked a new component to explain the hard \textit{Chandra} spectrum: cumulative SR from pair plasma in MSP magnetospheres. The low-energy tail of this HESR component reproduces the spectral slope of the X-ray data quite well. We argued that the required energetics and numbers of the MSP source population needed to reproduce the detected diffuse X-ray emission are plausible, albeit not very well constrained (although X-ray efficiencies of $\eta_{\rm X}\sim1\%$ and thus $\gamma_{\rm L}\sim0.7$ and $N^{\rm X}_{\rm vis}\sim10$ may be preferable). The MSP scenario to explain the broadband SED of \ter should thus be further scrutinized by future constraints on the properties (e.g., number of visible X-ray pulsars and their average spin-down luminosity) of the MSPs embedded in this GC.

For the VHE band, there were no new data available. The high-energy tail of our predicted unpulsed IC component produced a good fit to the current H.E.S.S\ data. More data obtained by  H.E.S.S.\ or new data from the CTA may better constrain the shape and cutoffs of the IC component, owing to the lower energy threshold as well as increased sensitivity of the latter. This may limit the particle minimum and maximum energies, source strength, average multiplicity, as well as the conversion efficiency of spin-down luminosity to particle power.

We have modelled the pulsed SR and CR components using a magnetospheric pulsar model, while we have modelled the LESR and IC components using an independent transport and emission model. While we have attempted to apply both these codes simultaneously for consistent parameter choices, a unified approach may lead to even deeper constraints on the cluster environment and stellar members. 

The morphology of structures associated with \ter differ significantly in extent and position for the different energy bands, challenging the idea that single particle population is responsible for all spectral components. Higher-resolution images of the GC will aid in elucidating the spatial properties of the different emission structures, possibly constraining the diffusion coefficient and cluster $B$-field profile. 

Using \ter as a case study, we could constrain our leptonic model for broadband emission from GCs. CTA will probably detect many more VHE GCs \citep{Ndiyavala18}, while multi-wavelength data on these sources should also continue to improve in both quantity and quality. This will allow us to further scrutinize competing emission models, as well as developing new, more complete and comprehensive ones that might explain the spatial \textit{and} spectral properties of Galactic GCs at an ever increasing level of detail.

\acknowledgments
We acknowledge fruitful discussions with Michael Backes, Carlo van Rensburg, Matthew Kerr, Hongjun An, Daniel Castro, Gudlaugur Johannesson, and Philippe Bruel. This work is based on the research supported wholly / in part by the National Research Foundation of South Africa (NRF; Grant Numbers 81671, 87613, 90822, 92860, 93278, and 99072). The Grantholder acknowledges that opinions, findings and conclusions or recommendations expressed in any publication generated by the NRF supported research is that of the author(s), and that the NRF accepts no liability whatsoever in this regard. A.K.H. acknowledges the support from the NASA Astrophysics Theory Program. C.V.\ and A.K.H.\ acknowledge support from the \textit{Fermi} Guest Investigator Program.

The \textit{Fermi} LAT Collaboration acknowledges generous ongoing support from a number of agencies and institutes that have supported both the development and the operation of the LAT as well as scientific data analysis. These include the National Aeronautics and Space Administration and the Department of Energy in the United States, the Commissariat \`a l'Energie Atomique and the Centre National de la Recherche Scientifique / Institut National de Physique Nucl\'eaire et de Physique des Particules in France, the Agenzia Spaziale Italiana and the Istituto Nazionale di Fisica Nucleare in Italy, the Ministry of Education, Culture, Sports, Science and Technology (MEXT), High Energy Accelerator Research Organization (KEK) and Japan Aerospace Exploration Agency (JAXA) in Japan, and
the K.~A.~Wallenberg Foundation, the Swedish Research Council and the Swedish National Space Board in Sweden.
 
Additional support for science analysis during the operations phase is gratefully acknowledged from the Istituto Nazionale di Astrofisica in Italy and the Centre National d'\'Etudes Spatiales in France. This work performed in part under DOE Contract DE-AC02-76SF00515.

\vspace{5mm}
\facilities{\textit{Fermi} Large Area Telescope, Effelsberg Radio Telescope, \textit{Chandra} X-ray Satellite, H.E.S.S.}


\begin{thebibliography}{}
\bibitem[Abdo et al.(2009)]{Abdo09} Abdo, A.~A., Ackermann, M., Ajello, M.\ et al. 2009, Science, 325, 845
\bibitem[Abdo et al.(2010)]{Abdo10} Abdo, A.~A., Ackermann, M., Ajello, M.\ et al. 2010, \aap, 524, A75
\bibitem[Abdo et al.(2013)]{2PC} Abdo, A.~A., Ajello, M., Allafort, A.~et al. 2013, \apjs, 208, 17
\bibitem[Abramowski et al.(2011a)]{Abramowski11_DM} Abramowski, A., Acero, F., Aharonian, F. et al. 2011a, \apj, 735, 12
\bibitem[Abramowski et al.(2011b)]{Abramowski11} Abramowski, A., Acero, F., Aharonian, F. et al. 2011b, \aap, 531, L18
\bibitem[Abramowski et al.(2013)]{Abramowski13} Abramowski, A., Acero, F., Aharonian, F.\ et al. 2013, \aap, 551, A26
\bibitem[Acero et al.(2015)]{3FGL} Acero, F., Ackermann, M., Ajello, M.\ et al. 2015, \apjs, 218, 23 
\bibitem[Acero et al.(2016)]{Acero16} Acero, F., Ackermann, M., Ajello, M.\ et al. 2016, \apjs, 223, 26 
\bibitem[Aharonian et al.(2009)]{Aharonian09} Aharonian, F. Akhperjanian, A.~G., Anton, G., et al. 2009, \aap, 499, 273
\bibitem[Anderhub et al.(2009)]{Anderhub09} Anderhub, H., Antonelli, L.~A., Antoranz, P. et al. 2009, \apj, 702, 266
\bibitem[Atoyan et al.(2006)]{Atoyan06} Atoyan, A., Buckley, J. \& Krawczynski, H. 2006, \apj, 642, L153
\bibitem[Atwood et al.(2009)]{Atwood09} Atwood, W.~B., Abdo, A.~A., Ackermann, M. et al. 2009, \apj, 697, 1071
\bibitem[Atwood et al.(2013)]{AtwoodP8} Atwood, W.~B., Albert, A., Baldini, L.\ et al. 2013, Proc.~of the 4th International Fermi Symposium, eConf C121028 (arXiv:1303.3514)
\bibitem[Barnard et al.(2016)]{Barnard16} Barnard, M., Venter, C., \& Harding, A.~K. 2016, \apj, 832, 107
\bibitem[Bednarek \& Sitarek(2007)]{BS07} Bednarek, W., Sitarek, J. 2007, \mnras, 377, 920 
\bibitem[Bednarek(2011)]{Bednarek11} Bednarek, W. 2011, in: D.~F.\ Torres and N.\ Rea (Eds.), High-energy emission from pulsars and their systems, 85
\bibitem[Bednarek(2012)]{Bednarek12} Bednarek, W. 2012, J. Phys. G Nucl. Phys., 39, 065001
\bibitem[Bednarek \& Sobczak(2014)]{Bednarek14} Bednarek, W., \& Sobczak, T. 2014, \mnras, 445, 2842
\bibitem[Bednarek et al.(2016)]{Bednarek16} Bednarek, W., Sitarek, J., \& Sobczak, T. 2016, \mnras, 458, 1083
\bibitem[Bhatia et al.(1992)]{Bhatia92} Bhatia, V.~B., Mishra, S., \& Panchapakesan, N. 1992, J.\ Astrophys.\ Astron., 13, 287
\bibitem[Bogdanov(2008)]{Bogdanov08} Bogdanov, S., Grindlay, J.~E., Rybicki, G.~B. 2008, \apj, 689, 407 
\bibitem[Brown et al.(2018)]{Brown18} Brown, A.~M., Lacroix, T., Lloyd, S., B{\oe}hm, C., \& Chadwick, P. 2018, \prd, 98, 041301
\bibitem[B\"usching et al.(2008)]{Buesching08} B{\"u}sching, I., de Jager, O.~C., Potgieter, M.~S., \& Venter, C. 2008, \apjl, 678, L39
\bibitem[Cadelano et al.(2018)]{Cadelano18} Cadelano, M., Ransom, S.~M., Freire, P.~C.~C., Ferraro, F.~R., Hessels, J.~W.~T., Lanzoni, B., Pallanca, C., \& Stairs I.~H. 2018, \apj, 855, 125
\bibitem[Camilo \& Rasio(2005)]{Camilo05} Camilo, F., \& Rasio, F.~A. 2005, ASPC, 328, 147
\bibitem[Chen(1991)]{Chen1991} Chen, K. 1991, \nat, 352, 695
\bibitem[Cheng et al.(2010)]{Cheng10} Cheng, K.~S., Chernyshov, D.~O., Dogiel, V.~A., Hui, C.~Y. \& Kong, A.~K.~H. 2011, \apj, 723, 1219 
\bibitem[Clapson et al.(2011)]{Clapson2011} Clapson, A.-C., Domainko, W., Jamrozy, M., Dyrda, M., Eger, P., 2011, \aap, 532, 47
\bibitem[Cohn et al.(2002)]{Cohn2002} Cohn, H.~N., Lugger, P.~M., Grindlay, J.~E., \& Edmonds, P.~D. 2002, \apj, 571, 818
\bibitem[Condon et al.(1998)]{Condon98} Condon, J.~J., Cotton, W.~D., Greisen, E.~W., Yin, Q.~F., Perley, R.~A., Taylor, G.~B., \& Broderick, J.~J. 1998, \aj, 115, 1693
\bibitem[de Jager et al.(1989)]{deJager89} de Jager, O.~C., Raubenheimer, B.~C., \& Swanepoel, J.~W.~H. 1989, \aap, 221, 180
\bibitem[de Jager \& B{\"u}sching(2010)]{deJager10} de Jager, O.~C. \& B{\"u}sching 2010, \aap, 517, L9
\bibitem[de Menezes et al.(2018)]{deMenezes} de Menezes, R., Cafardo, F., \& Nemmen, R. 2018, submitted to MNRAS (arXiv:1811.06957)
\bibitem[Domainko(2011)]{Domainko11} Domainko, W.~F., 2011, \aap, 533, L5
\bibitem[Dyks et al.(2004)]{Dyks04} Dyks, J. \& Harding, A.~K. 2004, \apj, 614, 869
\bibitem[Eger et al.(2010)]{Eger10} Eger, P., Domainko, W., \& Clapson, A.-C. 2010, \aap, 513, A66
\bibitem[Eger \& Domainko(2012)]{Eger12} Eger, P., \& Domainko, W. 2012, \aap, 540, A17
\bibitem[Emerson et al.(1979)]{emerson1979} Emerson, D.~T., Klein, U., Haslam, C.~G.~T., 1979, \aap, 76, 92
\bibitem[Emerson \& Gr{\"a}ve(1988)]{emerson1988} Emerson, D. T., Gr{\"a}ve, R. 1988, \aap, 190, 353
\bibitem[Ferraro et al.(2009)]{Ferraro09} Ferraro, F.~R., Dalessandro, E., Mucciarelli, A.\ et al. 2009, \nat, 462, 483
\bibitem[Foreman-Mackey et al.(2013)]{EMCEE} Foreman-Mackey, D., Hogg, D.~W., Lang, D., \& Goodman, J. 2013, \pasp, 125, 306
\bibitem[Freire(2016)]{Freire16} Freire P.~C.~C., 2016, Available at: http://www.naic.edu/$\sim$pfreire/GCpsr.html
\bibitem[Freire et al.(2017)]{Freire17} Freire, P.~C.~C., Ridolfi, A., Kramer, M.\ et al. 2017, \mnras, 471, 857
\bibitem[Fruchter \& Goss(2000)]{Fruchter00} Fruchter, A.~S., \& Goss, W.~ M. 2000, \apj, 536, 865
\bibitem[Fukui et al.(2009)]{Fukui09} Fukui, Y., Furukawa, N., Dame, T.~M. et al. 2009, \pasj, 61, L23
\bibitem[Gioia et al.(1982)]{gioia1982} Gioia, I.M., Gregorini, L., Klein, U. 1982, \aap, 116, 164
\bibitem[Goldreich \& Julian(1969)]{GJ69} Goldreich, P., \& Julian, W.~H. 1969, \apj, 157, 869
\bibitem[Gonthier et al. (2018)]{Gonthier2018} Gonthier, P.~L., Harding, A.~K., Ferrara, E.~C., Frederick, S.~E., Mohr, V.~E. \& Koh, Y-M.  2018, \apj, 863, 199
\bibitem[Harding et al.(1978)]{Harding78} Harding, A.~K., Tademaru, E. \& Esposito, L.~W. 1978, \apj, 225, 226
\bibitem[Harding et al.(2005)]{HUM05} Harding, A.~K., Usov, V.~V., \& Muslimov, A.~G. 2005, \apj, 622, 531
\bibitem[Harding et al.(2008)]{Harding08} Harding, A.~K., Stern, J.~V., Dyks, J., \& Frackowiak, M. 2008, \apj, 680, 1378
\bibitem[Harding \& Muslimov(2011a)]{Harding11a} Harding, A.~K., \& Muslimov, A.~G. 2011a, \apjl, 726, L10
\bibitem[Harding \& Muslimov(2011b)]{Harding11b} Harding, A.~K., \& Muslimov, A.~G. 2011b, \apj, 743, 181
\bibitem[Harding \& Kalapotharakos(2015)]{HK15} Harding, A.~K., \& Kalapotharakos, C. 2015, \apj, 811, 63
\bibitem[Haslam(1974)]{haslam1974} Haslam, C.~G.~T. 1974, \aaps, 15, 333
\bibitem[Heinke et al.(2006)]{Heinke06} Heinke, C.~O., Wijnands, R., Cohn, H.~N., Lugger, P.~M., Grindlay, J.~E., Pooley, D., \& Lewin, W.~H.~G. 2006, \apj, 651, 1098
\bibitem[Hessels et al.(2006)]{Hessels06} Hessels, J.~W.~T., Ransom, S.~M., Stairs, I.~H., Freire, P.~C.~C., Kaspi, V.~M., \& Camilo, F. 2006, Science, 311, 1901
\bibitem[Ivanova et al.(2008)]{Ivanova08} Ivanova, N., Heinke, C.~O., \& Rasio, F.~A. 2008, in 40 Years of Pulsars: Millisecond Pulsars, Magnetars and More, AIP Conf.\ Ser., ed. C. Bassa, Z. Wang, Z., A. Cumming, \& V.~M. Kaspi, 983, 442 (arXiv:0711.3001)
\bibitem[Johnson et al.(2015)]{JohnsonJ1227} Johnson, T.~J., Ray, P.~S., Roy, J.~et al.~2015, \apj, 806, 91
\bibitem[Johnston \& Verbunt(1996)]{Johnston1996} Johnston, H.~M.\& Verbunt, F. 1996, \aap, 312, 80
\bibitem[Kalapotharakos et al.(2018)]{Kala18} Kalapotharakos, C., Brambilla, G., Timokhin, A., Harding, A.~K. \& Kazanas, D.   2018, \apj, 857, 44
\bibitem[Kerr(2011)]{Kerr11} Kerr, M.~2011, \apj, 732, 38
\bibitem[King(1962)]{King62} King, I.~R. 1962, \aj, 67, 471
\bibitem[Kong et al.(2010)]{Kong10} Kong, A.~K.~H., Hui, C.~Y. \& Cheng, K.~S. 2010, \apj, 712, 36
\bibitem[Kopp et al.(2013)]{Kopp13} Kopp, A., Venter, C., B\"usching, I., \& de Jager, O.~C. 2013, \apj, 779, 126
\bibitem[Lang(1992)]{Lang1992} Lang, K.~R. 1992, Astrophysical Data I. Planets and Stars, Berlin: Springer-Verlag
\bibitem[Lanzoni et al.(2011)]{Lanzoni10} Lanzoni, B., Ferraro, F.~R., Dalessandro, E.\ et al. 2010, \apj, 717, 653
\bibitem[Leung et al.(2014)]{Leung14} Leung, G.~C.~K., Takata, J., Ng, C.~W. et al. 2014, \apjl, 797, L13
\bibitem[Lyne et al.(1990)]{Lyne90} Lyne, A.~G., Johnston, S., Manchester, R.~N., Staveley-Smith, L., \& D'Amico, N. 1990, \nat, 347, 650
\bibitem[Lyne et al.(2000)]{Lyne00} Lyne, A.~G., Mankelow, S.~H., Bell, J.~F., \& Manchester, R.~N. 2000, \mnras, 316, 491
\bibitem[McCutcheon et al.(2009)]{McCutcheon09} McCutcheon M et al. 2009 {\it Proc. 31st ICRC, Lodz, Poland} (arXiv:0907.4974)
\bibitem[Morsi \& Reich(1988)]{morsi1986} Morsi, H.W., Reich, W. 1986, \aap, 163, 313\
\bibitem[Ndiyavala et al.(2018)]{Ndiyavala18} Ndiyavala, H., Kr{\"u}ger, P.~P., \& Venter, C. 2018, \mnras, 473, 897 
\bibitem[Nolan et al.(2012)]{2FGL} Nolan, P.L., Abdo, A.~A., Ackermann, M.\ et al. 2012, \apjs, 199, 31
\bibitem[Prinsloo et al.(2013)]{Prinsloo13} Prinsloo, P., Venter, C., B{\"u}sching, I.,  \& Kopp, A, 2013, arXiv:1311.3791
\bibitem[Ransom et al.(2005)]{Ransom05} 
Ransom, S.~M., Hessels, J.~W.~T., Stairs, I.~H., Freire, P.~C.~C., Camilo, F., Kaspi, V.~M., \& Kaplan, D.~L. 2005, Science, 307, 892
\bibitem[Roy et al.(2005)]{roy2005} Roy, S., Rao, A.P., Subrahmanyan, R., 2005, \mnras, 360, 1305
\bibitem[Schlickeiser \& Ruppel(2010)]{Ruppel10} Schlickeiser, R., \& Ruppel, J. 2010, New J. Phys., 12, 3
\bibitem[Tam et al.(2016)]{Tam16} Tam, P.-H.~T., Hui, C.~Y., \& Kong, A.~K.~H. 2016, J.\ Astron.\ SS., 33, 1
\bibitem[Tasker et al.(1994)]{tasker1994} Tasker, N.J., Condon, J.J., Wright, A.E., Griffith, M.R. 1994, \aj, 107, 2115
\bibitem[The \textit{Fermi}-LAT collaboration.(2019)]{4FGL} The \textit{Fermi}-LAT collaboration, 2019, arXiv:1902.10045
\bibitem[Trager et al.(1995)]{Trager1995} Trager, S.~C., King, I.~R., \& Djorgovski, S. 1995, \aj, 109, 218
\bibitem[Valenti et al.(2007)]{Valenti07} Valenti, E., Ferraro, F.~R., \& Origlia, L. 2007, \apj, 133, 1287
\bibitem[Venter \& de Jager(2008)]{Venter08} Venter, C., \& de Jager, O.~C. 2008, AIP Conf.\ Ser., 1085, 277
\bibitem[Venter et al.(2009a)]{Venter09_GC} Venter, C., de Jager, O.~C., \& Clapson, A.-C. 2009a, \apj, 696, L52
\bibitem[Venter \& Kopp(2015a)]{Venter15} Venter, C., \& Kopp, A. 2015a, Proc.\ SAIP2014, ed.\ Chris Engelbrecht \& Steven Karataglidis, 394 (arXiv:1504.04953)
\bibitem[Venter \& Kopp(2015b)]{Venter_HEASA15} Venter, C., \& Kopp., A. 2015b, in: High-energy Astrophysics in Southern Africa 2014: A Multi-frequency Perspective of New Frontiers in High-energy Astrophysics in Southern Africa, Proc.\ HEASA2014, Mem.\ Soc.\ Astron.\ It., ed.\ B.\ van Soelen and P.~J.\ Meintjes, 86,70
\bibitem[Venter et al.(2015c)]{Venter15c} Venter, C., \& Kopp, A., Harding, A.~K., Gonthier, P.~L., \& B\"usching, I. 2015c, \apj, 807, 130
\bibitem[Verbunt \& Hut(1987)]{Verbunt87} Verbunt, F., \& Hut, P. 1987, in IAU Symp.\ 125, The Origin and Evolution of Neutron Stars, ed.\ D.~J.\ Helfand \& H.~H.\
Huang (Dordrecht: Reidel), 187
\bibitem[Wei et al.(1996)]{Wei96} Wei, D.~M., Cheng, K.~S., \& Lu T. 1996, \apj, 468, 207
\bibitem[Wu et al.(2014)]{Wu14} Wu, E.~M.~H., Hui, C.~Y., Kong, A.~K.~H., Tam, P.~H.~T., Cheng, K.~S., \& Dogiel, V.~A., 2014, \apjl, 788, L40
\bibitem[Yamamoto et al.(2008)]{Yamamoto08} Yamamoto, H. et al. 2008, PASJ, 60, 715
\bibitem[Zajczyk et al.(2013)]{Zajczyk13} Zajczyk, A., Bednarek, W. \& Rudak, B. 2013, \mnras, 432, 3462
\bibitem[Zhang et al.(2016)]{Zhang16} Zhang, P.~F., Xin, Y.~L., Fu, L., Zhou, J.~N., Yan, J.~Z., Liu, Q.~Z., \& Zhang, L. 2016, \mnras, 459, 99
\end{thebibliography}
\end{document}